\documentclass[prd,twocolumn,showpacs,superscriptaddress,nofootinbib,floatfix,showkeys,10pt]{revtex4-2}

\usepackage{graphicx}
\usepackage{amsmath}
\usepackage{bm}
\usepackage{yhmath}
\usepackage{mathtools}
\usepackage{wasysym}
\usepackage[colorlinks,citecolor=blue,urlcolor=blue,linkcolor=blue]{hyperref}
\usepackage{subfigure}
\usepackage{color}
\usepackage{cases}
\usepackage{times}
\usepackage{dcolumn,booktabs,bm}
\usepackage{slashed}
\usepackage{amsfonts,amssymb,stmaryrd,latexsym,amsmath}
\usepackage{textcomp}
\usepackage{multirow}
\usepackage{cancel}
\usepackage{array}
\usepackage{orcidlink}
\usepackage{enumitem}
\usepackage{dashrule}

\allowdisplaybreaks[1]

\renewcommand{\arraystretch}{1.8}

\begin{document}

\title{Bound and Resonant Spectra of Few-Lepton Coulomb Systems}

\author{Liang-Zhen Wen\,\orcidlink{0009-0006-8266-5840}}\email{wenlzh\_hep-th@stu.pku.edu.cn}
\affiliation{School of Physics, Peking University, Beijing 100871, China}

\author{Yao Ma\,\orcidlink{0000-0002-5868-1166}}\email{yao.ma@tum.de}
\affiliation{Technical University of Munich, TUM School of Natural Sciences, Physics Department, James-Franck-Str. 1, 85748 Garching, Germany}

\author{Zhi-He Shen\,\orcidlink{0009-0009-6836-9780}}\email{papercrane@stu.pku.edu.cn}
\affiliation{School of Physics, Peking University, Beijing 100871, China}

\author{Shi-Lin Zhu\,\orcidlink{0000-0002-4055-6906}}\email{zhusl@pku.edu.cn}
\affiliation{School of Physics and Center of High Energy Physics, Peking University, Beijing 100871, China}

\begin{abstract}
We present a unified calculation of bound and resonant states in purely
leptonic Coulomb systems: $e^-e^-e^+$ ($\mathrm{Ps}^-$),
$\mu^+e^-e^-$ ($\mathrm{Mu}^-$), $\mu^+\mu^+e^-$ ($\mathrm{Mu}_2^+$),
$e^+e^+e^-e^-$ ($\mathrm{Ps}_2$), and $\mu^+\mu^+e^-e^-$
($\mathrm{Mu}_2$). Using an extended
stochastic variational method combined with the complex scaling method, we resolve the natural-parity $S$- and $P$-wave spectra.
Near their respective $n=2$ thresholds,
all three trilepton systems exhibit Gailitis--Damburg sequences generated by
$2S$ and $2P$ Stark mixing and the resulting inverse-square attraction.
Although microscopically distinct from the Efimov effect, this mechanism produces
the same inverse-square asymptotics and geometric scaling.
 The $\mathrm{Ps}^-$ and $\mathrm{Mu}^-$ systems exhibit resonance sequences
 of comparable density, whereas the $\mathrm{Mu}_2^+$
produces a much denser spectrum, with 20 resolved members in the $^3P^o$ channel.
In $\mathrm{Mu}_2^+$, the deeper states follow molecular Born--Oppenheimer configurations,
while the near-threshold spectrum is governed by the atomic $\mathrm{Mu}(2)+\mu^+$ structure.
In $\mathrm{Ps}_2$, coupling between threshold-degenerate configurations is
essential for a near-threshold bound state. In $\mathrm{Mu}_2$, the
Born--Oppenheimer organization of the resonance spectrum is channel dependent.
\end{abstract}

\maketitle

\section{Introduction}\label{sec:intro}

Few-lepton systems are among the simplest composite systems governed by the QED
Coulomb interaction. They consist only of charged leptons, so neither the
strong interaction nor nuclear structure plays a role. Following Wheeler's proposal of polyelectrons~\cite{Wheeler:1946xth},
the positronium negative ion $e^-e^-e^+$ ($\mathrm{Ps}^-$) was found to
support a bound ground state
~\cite{frost1964approximate,Martin:1991ca,Ho:1993zz,drake2002ground}.
The positronium molecule $e^+e^+e^-e^-$ ($\mathrm{Ps}_2$) also supports a
bound ground state and excited states
~\cite{Hylleraas:1947zza,ho1990positronium,Varga:1998ss,
suzuki2000excited,Puchalski:2008jj,matyus2012molecular}.
Both $\mathrm{Ps}^-$ and $\mathrm{Ps}_2$ have been observed experimentally~\cite{Mills:1981zzc,Cassidy:2007blx}.
An $L=1$ excited state of $\mathrm{Ps}_2$ has also been probed using optical spectroscopy~\cite{Cassidy_2012}.
Beyond the bound-state sector, the resonance spectra of $\mathrm{Ps}^-$~\cite{Ho:1979zz,ho1984doubly,rost1992positronium,ivanov1999high,Usukura:2002zz,suzuki2004stochastic,kar2009d,ho2012complex,matyus2013resonances,igarashi2016broad,Kar:2018sax,Kar:2018vmc,bhatia1990pwave,kar2019calculations,kar2020triplet} and $\mathrm{Ps}_2$~\cite{ho1989resonant,Usukura:2002zz,bao2003lowest,suzuki2004stochastic,dirienzi2010resonances,matyus2013resonances,zhang2020doubly} have also been investigated theoretically.
A muon has the same electric charge and spin as an electron but is approximately 207 times heavier.
Thus muon-containing few-lepton systems realize the same Coulomb interaction in a strong mass imbalance regime, where distinct spatial structures can emerge. Previous studies of $\mu^+e^-e^-$ ($\mathrm{Mu}^-$), $\mu^+\mu^+e^-$ ($\mathrm{Mu}_2^+$), and $\mu^+\mu^+e^-e^-$ ($\mathrm{Mu}_2$) have mainly addressed bound states~\cite{Martin:1991ca,frolov1999bound,Frolov:2017tvw}, while resonance calculations have been limited to selected $S$-wave channels~\cite{Ho:1979zz,liverts2013three,Ma:2025rvj}. Developments in high-intensity muon facilities may bring more exotic leptonic systems within experimental reach~\cite{Achasov:2023gey,Prokscha:2008zz,Kanda:2023gqp,Bai:2024skk,Chen:2026tdg,Cai:2023caf,An:2025lws,Liu:2025ejy}.
We therefore extend these studies beyond the $S$ wave and systematically calculate the $P$-wave bound and resonant states.

The Efimov effect is a hallmark of universality in three-body physics~\cite{Efimov:1970zz,Braaten:2004rn,Naidon:2016dpf}. At unitarity, resonant short-range interactions generate
an attractive $-1/R^2$ hyperradial potential.
This potential supports an infinite geometric tower of three-body states.
 A related geometric structure arises in Coulomb systems near degenerate thresholds.
 Stark mixing between the $2S$ and $2P$ states produces an charge--dipole interaction proportional to $-1/R^2$.
 When this attraction is supercritical, it generates a Gailitis--Damburg sequence that
 accumulates at the threshold~\cite{gailitis1963,gailitis1982finite}.
 The two mechanisms have different microscopic origins but share inverse square asymptotics
 and discrete scale invariance.
The present work examines this charge--dipole organization in $\mathrm{Ps}^-$, $\mathrm{Mu}^-$, and $\mathrm{Mu}_2^+$.
Their resonace spectra form Gailitis--Damburg sequences, whose geometric scaling varies with the mass configuration.
In $\mathrm{Mu}_2^+$, The deeper spectrum exhibits a molecular Born--Oppenheimer (BO) structure. Near threshold, the atomic $\mathrm{Mu}(2)+\mu^+$ structure becomes dominant.

We systematically calculate the natural-parity $S$- and $P$-wave channels of $\mathrm{Ps}^-$, $\mathrm{Mu}^-$, $\mathrm{Mu}_2^+$, $\mathrm{Ps}_2$, and $\mathrm{Mu}_2$.
Their shallow resonances require a simultaneous description of compact structures
within the two-body clusters and extended motion between asymptotic clusters.
We combine the complex scaling method (CSM) with an extended stochastic variational method (ESVM)~\cite{Wen:2025ehf}.
The explicitly correlated Gaussian basis is augmented by atom--spectator and atom--atom configurations.
Their intercluster length scales cover the relevant short- and long-distance regions.
This framework treats bound states and resonant states within the same footing.
 Within the nonrelativistic Coulomb Hamiltonian, the equal-mass muonic systems $\mu^+\mu^+\mu^-$ and $\mu^+\mu^+\mu^-\mu^-$ follow by exact mass scaling.

The paper is organized as follows. Section~\ref{sec:framework} introduces the
Hamiltonian, wave-function construction, CSM, and ESVM basis. Section~\ref{sec:results}
presents the trilepton spectra. Section~\ref{sec:GD} analyzes the
universal dynamics of trilepton systems at $n=2$ thresholds.
Section~\ref{subsec:4l} presents the tetralepton spectra, and
Sec.~\ref{sec:sum} summarizes the main results and their implications.

\section{Theoretical framework}\label{sec:framework}

\subsection{Hamiltonian}\label{subsec:Hamiltonian}

The nonrelativistic Hamiltonian of the $N$-body QED system reads
\begin{align}\label{eq:Hamiltonian}
H=\sum_i^N\frac{\boldsymbol{p}_i^2}{2m_i}+\sum_{i<j=1}^N V_{ij}\,,
\end{align}
where $m_i$ and $\boldsymbol{p}_i$ are the mass and momentum of particle $i$, and $V_{ij}$ is the Coulomb interaction,
\begin{align}\label{eq:Vij}
V_{ij} = \alpha\frac{Q_i Q_j}{r_{ij}},
\end{align}
with $Q_i$ the charge of particle $i$ (in units of $e$) and
$r_{ij}=|\boldsymbol{r}_i-\boldsymbol{r}_j|$. All constituents are charged
leptons, $e^\pm$ or $\mu^\pm$, so $Q_i=\pm1$. The lepton masses and the
fine-structure constant $\alpha$ are taken from CODATA~\cite{Mohr:2024kco} and
listed in Table~\ref{tab:mass}. The relevant two-body subsystems are the
positronium atom $e^+e^-$(Ps) and the muonium atom $\mu^+e^-$(Mu). In Mu, the
antimuon $\mu^+$ plays the role of the nucleus. Their energy levels are hydrogenic.
Computed here with the same method, they set dissociation thresholds and are
collected in Table~\ref{tab:twobody}.

\begin{table}[htbp]
    \centering
    \caption{The electron and muon masses (in MeV) and the fine-structure constant $\alpha$, from CODATA~\cite{Mohr:2024kco}.}
    \label{tab:mass}
    \begin{tabular*}{\hsize}{@{}@{\extracolsep{\fill}}ccc@{}}
\hline\hline
 $m_e$ & $m_\mu$ & $\alpha$ \\
\hline 0.51099895 & 105.65837 & $\tfrac{1}{137.03600}$ \\
\hline\hline
    \end{tabular*}
\end{table}

\begin{table}[htbp]
\renewcommand{\arraystretch}{1.4}
\centering
\caption{\label{tab:twobody} Energies $E_n$ and rms radii
$r^{\mathrm{rms}}_{nl}=\langle r^2\rangle^{1/2}$ of the two-body Ps and Mu
atoms, computed with the present method. These levels define the dissociation
thresholds used in the three- and four-body calculations. The $2S$ and $2P$
levels are energy-degenerate but have different radii.}
\begin{tabular*}{\hsize}{@{}@{\extracolsep{\fill}}lcc@{}}
\hline\hline
  & $E\,[\mathrm{eV}]$ & $r^{\mathrm{rms}}\,[\mathrm{nm}]$ \\
\hline
 $\mathrm{Ps}(1S)$ & $-6.8028$  & $0.183$ \\
 $\mathrm{Ps}(2S)$ & $-1.7007$  & $0.686$ \\
 $\mathrm{Ps}(2P)$ & $-1.7007$  & $0.580$ \\
 $\mathrm{Mu}(1S)$ & $-13.5402$ & $0.092$ \\
 $\mathrm{Mu}(2S)$ & $-3.3851$  & $0.345$ \\
 $\mathrm{Mu}(2P)$ & $-3.3851$  & $0.291$ \\
\hline\hline
\end{tabular*}
\end{table}

\subsection{Wave function construction}\label{subsec:wavefunction}

The wave function is a product of the spatial part $\chi_r$ and the spin part $\chi_s$, projected onto the physical symmetry,
\begin{equation}\label{eq:Abasis}
\psi = \mathcal{A}\left( \chi_{r} \otimes \chi_{s} \right),
\end{equation}
where $\mathcal{A}$ is the (anti-)symmetrization operator and all constituents are fermions ($e^\pm,\mu^\pm$). Since the interaction is purely Coulombic, the total orbital angular momentum $L$ is conserved, and the total spin of each identical like-sign pair is a good quantum number, denoted $S_{12}$ for the pair $(1,2)$ and $S_{34}$ for the pair $(3,4)$.

For the trilepton systems with two identical leptons (e.g.\ $e^-e^-e^+$, $\mu^+\mu^+e^-$), only particles $l_1^\pm,l_2^\pm$ are identical, so $\mathcal{A}=1-P_{12}$, and the spin function couples them into a pair of spin $S_{12}$,
\begin{equation}\label{eq:spin3}
(l_1^\pm l_2^\pm)_{S_{12}},\qquad S_{12}=0,1,
\end{equation}
which enforces the correct (anti-)symmetry of the spatial wave function (the spin of the third lepton plays no role). Each channel is thus specified by the two quantum numbers $(S_{12},L)$, which are used to label the channels in the trilepton figures. Equivalently, we use the spectroscopic term $^{2S_{12}+1}L^{\pi}$. Here $\pi$ denotes the spatial parity under inversion of all relative (Jacobi) coordinates, with the superscripts $e$ and $o$ indicating even and odd parity, respectively. In this work we consider only the natural-parity states, $\pi=(-1)^{L}$, i.e.\ $^1S^e$, $^3S^e$, $^1P^o$, and $^3P^o$ for the $S$- and $P$-wave channels studied here.

\begin{figure}[tb]
\centering
\includegraphics[width=0.24\textwidth]{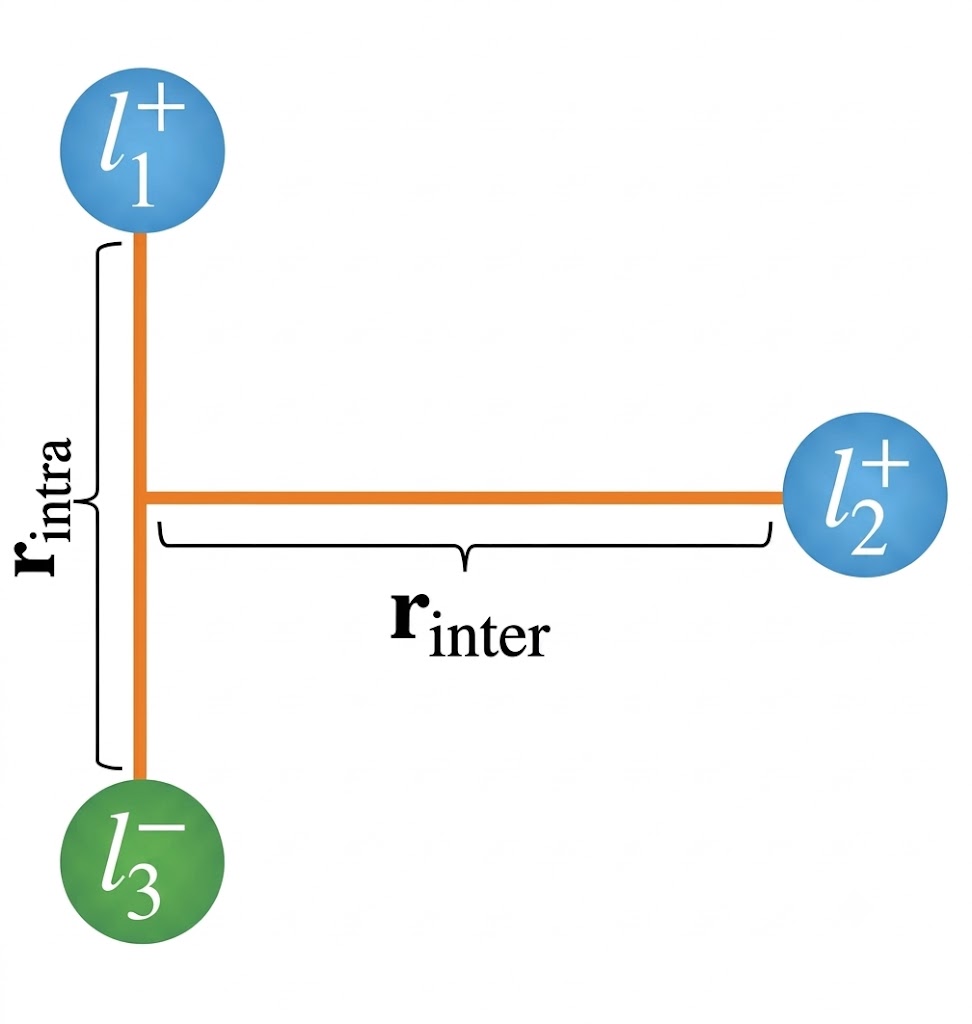}
\caption{\label{fig:structure_3l} Molecular configuration used in the trilepton
basis. A bound atom, shown here as $l_1^+l_3^-$, is described by
$\boldsymbol{r}_{\rm intra}$, and the remaining lepton $l_2^+$ is described by
$\boldsymbol{r}_{\rm inter}$.}
\end{figure}

For the tetralepton systems $e^+e^+e^-e^-$, $\mu^+\mu^+\mu^-\mu^-$, and $\mu^+\mu^+e^-e^-$, both the like-sign positive pair and the negative pair are identical fermions, so
\begin{equation}
\mathcal{A}=(1-P_{12})(1-P_{34}),
\end{equation}
and the spin function couples each pair separately,
\begin{equation}\label{eq:spin4}
(l_1^+ l_2^+)_{S_{12}}\,(l_3^- l_4^-)_{S_{34}},\qquad S_{12},S_{34}=0,1,
\end{equation}
where both pair spins, $S_{12}$ and $S_{34}$, remain good quantum numbers. We
therefore label each tetralepton channel by $(S_{12},S_{34},L)$. For example,
$(0,0,0)$ denotes
the doubly singlet $S$-wave channel.

For the spatial part, we use a basis of explicitly correlated Gaussians
(ECGs)~\cite{Mitroy:2013eom},
\begin{equation}\label{eq:basisSpace}
\psi(\mathbf r)=\theta_{KLM}(\mathbf{v})\exp\!\left(-\sum_{i>j=1}^N \alpha_{ij}(\mathbf{r}_i-\mathbf{r}_j)^2\right),
\end{equation}
with nonlinear parameters $\alpha_{ij}$. Higher orbital angular momenta are incorporated through the global vector representation (GVR)~\cite{Suzuki:1997tu},
\begin{equation}
\theta_{KLM}(\mathbf{v})=|\mathbf{v}|^{2K+L}\,Y_{LM}(\hat{\mathbf{v}}),
\qquad \mathbf{v}=\sum_{i=1}^{N}u_i\,\mathbf{r}_i ,
\end{equation}
where the coefficients $u_i$ select the geometric configuration carrying the angular momentum, and the center-of-mass redundancy is removed by imposing $\sum_i u_i=0$. The integer $K\ge0$ adds flexibility to the short-range behavior.

A key advantage of the GVR is that all overlap and Hamiltonian matrix elements can be evaluated in closed form. The explicit formulas are given in Ref.~\cite{Suzuki:1997tu}.

\subsection{Complex scaling method}\label{subsec:method}

\begin{figure}[tbp]
\centering
\includegraphics[width=0.9\columnwidth]{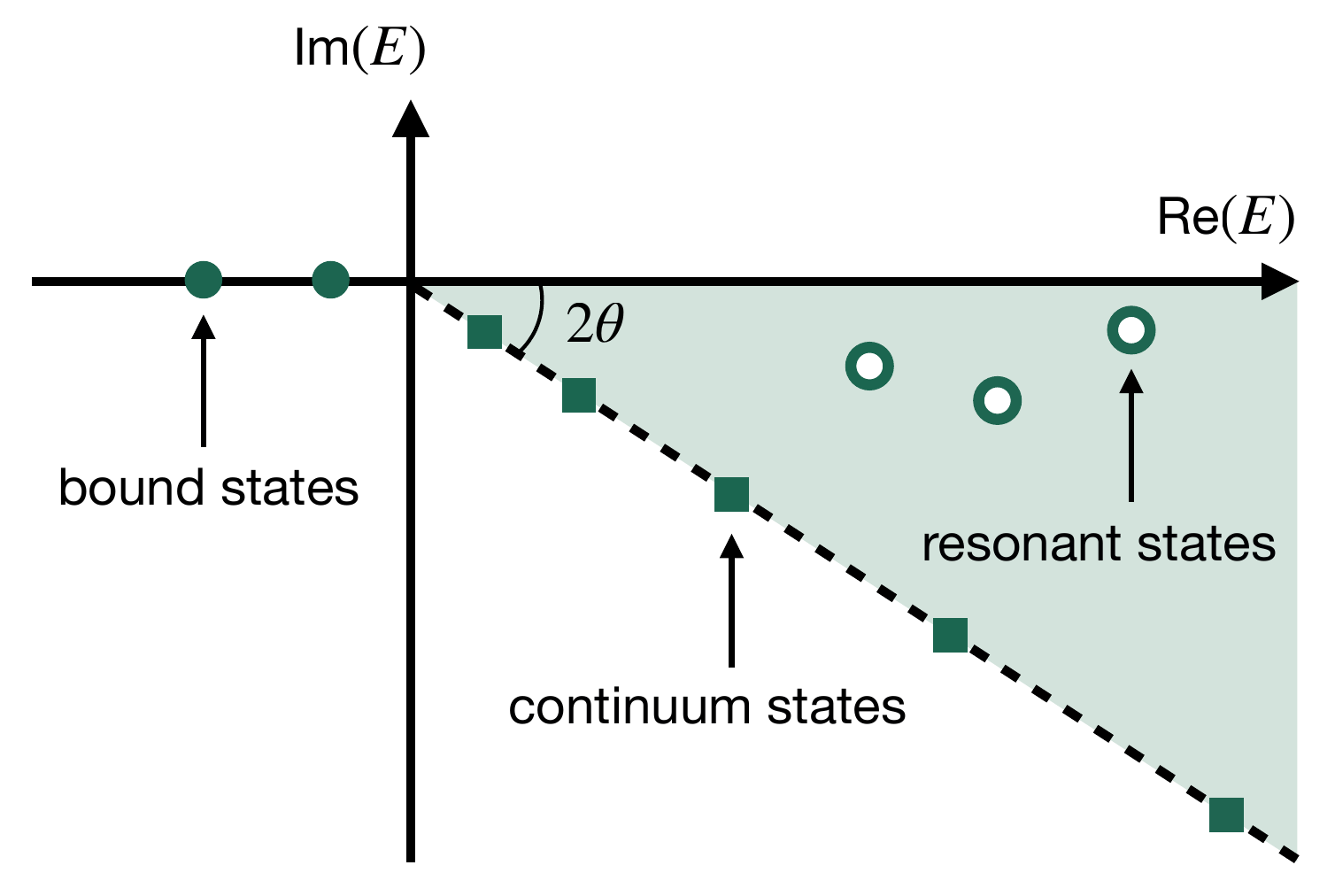}
\caption{\label{fig:csm} Schematic eigenvalue distribution of the complex-scaled
Hamiltonian $H(\theta)$ for a two-body system. Bound states (solid circles) lie
on the negative real axis. The continuum of each channel (squares) rotates onto
the line $\operatorname{Arg}(E)=-2\theta$ from its threshold. Resonant states
(open circles) are uncovered at $E_{\rm res}=E_r-i\Gamma/2$ and stay fixed as $\theta$
is varied. Reproduced from Ref.~\cite{Ma:2025rvj}.
\href{https://creativecommons.org/licenses/by/4.0/}{CC BY 4.0}.}
\end{figure}

Bound and resonant states are obtained on the same footing with the
CSM~\cite{Aguilar:1971ve,Balslev:1971vb,Aoyama:2006hrz}. The method gives the
resonance energies and widths directly through an analytic continuation of the
Schr\"odinger equation. The coordinates and momenta are rotated,
\begin{align}\label{eq:complexRotation}
U(\theta)\boldsymbol{r}=\boldsymbol{r}e^{i\theta},\qquad U(\theta)\boldsymbol{p}=\boldsymbol{p}e^{-i\theta},
\end{align}
under which the Hamiltonian of Eq.~\eqref{eq:Hamiltonian} becomes
\begin{equation}\label{eq:HamiltonianComplex}
H(\theta)=\sum_{i=1}^N\frac{\mathbf{p}_i^2 e^{-2i\theta}}{2m_i}+\sum_{i<j=1}^N V_{ij}\!\left(\mathbf{r}_{ij}e^{i\theta}\right).
\end{equation}
For a resonance whose pole lies within the rotation angle, the wave function
becomes square-integrable and can be expanded in the same localized basis as a
bound state. Diagonalizing $H(\theta)$ then yields bound and resonant states
simultaneously. In the complex energy plane, bound states lie on the negative
real axis. The continuum of each channel rotates into a cut with
$\operatorname{Arg}(E)=-2\theta$ from its threshold, while a resonance appears at
$E_{\rm res}=E_r-i\Gamma/2$ provided
$|\operatorname{Arg}(E_{\rm res})|<2\theta$. The positions of bound and resonant
states do not move as $\theta$ is varied. This $\theta$-stability distinguishes
genuine states from the rotated continuum. Fig.~\ref{fig:csm} shows the
resulting pattern schematically.

\subsection{Extended stochastic variational method}\label{subsec:ESVM}

We compute the bound and resonant states with the ESVM~\cite{Wen:2025ehf}. The trial wave function is a superposition of a random and a molecular basis,
\begin{equation}
\Psi_{\mathrm{trial}}=\Psi_{\mathrm{random}}+\Psi_{\mathrm{molecular}}.
\end{equation}

\begin{figure*}[t]
\centering
\includegraphics[width=\textwidth]{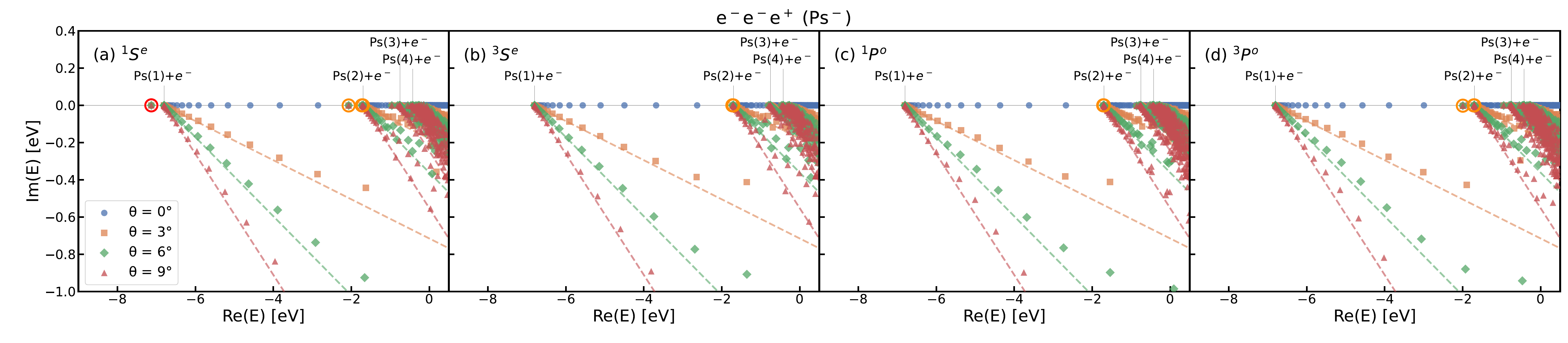}
\caption{\label{fig:eee} Complex-energy spectra of the $e^-e^-e^+$
system under complex scaling for the four natural-parity
channels (a)~$^1S^e$, (b)~$^3S^e$, (c)~$^1P^o$, and (d)~$^3P^o$. Continuum
eigenvalues align along $\mathrm{Arg}(E)=-2\theta$. Stable bound and resonant
poles are circled in red and orange, respectively.}
\end{figure*}

The random part $\Psi_{\mathrm{random}}$ is a system-dependent pool of explicitly
correlated Gaussians [Eq.~\eqref{eq:basisSpace}]. The nonlinear parameters
$\alpha_{ij}$ are sampled log-uniformly, with their ranges determined by the
radial windows listed in Appendix~\ref{app:Gaussian_par}. This set is generated
by the initial stochastic sampling step of the stochastic variational method
(SVM)~\cite{Varga:1995dm}. The usual growth and refinement cycles are not applied to the
random basis, because optimizing a single bound-state energy is not
well suited to locating resonant poles. Instead, we add the molecular part
$\Psi_{\mathrm{molecular}}$, a Gaussian basis built on the GEM~\cite{Hiyama:2003cu}
that makes the two-cluster structure explicit.
For a trilepton system, each molecular basis function is a product
\begin{equation}\label{eq:scatbasis}
\Psi_{\mathrm{molecular}}=\underbrace{\phi_{\rm atom}(\boldsymbol{r}_{\rm intra})}_{\text{bound atom}}\;
\underbrace{e^{-\boldsymbol{r}_{\rm inter}^2/r_n^2}}_{\text{relative motion}}\;
\underbrace{\theta_{LM}\!\big(\textstyle\sum_i u_i\boldsymbol r_i\big)}_{\text{global vector}}.
\end{equation}
Here $\phi_{\rm atom}$ describes the two-body cluster (Ps or Mu). It is a
Gaussian expansion optimized by the SVM and converged up to $n=4$ to capture the
near-threshold poles. The coordinate $\boldsymbol{r}_{\rm intra}$ is the
internal coordinate of the bound atom, and $\boldsymbol{r}_{\rm inter}$ is the
relative coordinate between the atom and the remaining lepton(s)
(Fig.~\ref{fig:structure_3l}). For tetralepton systems, the configuration
contains two atomic clusters. It therefore carries two $\phi_{\rm atom}$ factors,
and $\boldsymbol{r}_{\rm inter}$ is the relative coordinate between the two
atoms. The inter-cluster widths $r_n$ form a geometric progression,
\begin{equation}\label{eq:geometric}
r_n=r_1\,a^{\,n-1},\qquad a=\left(\frac{r_{n_{\max}}}{r_1}\right)^{\frac{1}{n_{\max}-1}},
\end{equation}
where the common ratio $a$ is fixed by the smallest and largest widths $r_1$ and $r_{n_{\max}}$ together with the number of terms $n_{\max}$. For a given $L$, there is a formal connection between the global-vector representation and the partial-wave expansion~\cite{Suzuki:1997tu}. For a global vector $\mathbf{v}=u_1\mathbf{x}_1+u_2\mathbf{x}_2$ built from two Jacobi vectors (shown for brevity),
\begin{widetext}
\begin{equation}\label{eq:addition}
\theta_{LM}(\mathbf{v})=|\mathbf{v}|^{L}Y_{LM}(\hat{\mathbf{v}})=\!\!\sum_{l_1+l_2=L}\!\!c^{LM}_{l_1 l_2}\,u_1^{l_1}u_2^{l_2}\,|\mathbf{x}_1|^{l_1}|\mathbf{x}_2|^{l_2}\,\big[Y_{l_1}(\hat{\mathbf{x}}_1)\otimes Y_{l_2}(\hat{\mathbf{x}}_2)\big]_{LM},
\end{equation}
\end{widetext}
a fixed combination of the coupled single-coordinate harmonics with coefficients
independent of $u_1,u_2$. This relation can be inverted. Each partial wave on
the right can be recovered as a linear combination of the left-hand side for
appropriate values of $u_1,u_2$, and the construction generalizes to $N$
particles. We therefore treat the $u_i$ as free parameters that generate the
orbital motion of each cluster configuration, while keeping all matrix elements
analytic. The molecular basis is essential for the shallow near-threshold
states, which a purely random basis cannot resolve. The per-system parameters
are listed in Appendix~\ref{app:Gaussian_par}.

\section{Numerical results of trilepton systems}\label{sec:results}

%\subsection{Trilepton systems}\label{subsec:3l}

\subsection{$e^-e^-e^+$ and $\mu^+\mu^+\mu^-$}\label{subsubsec:eee}

%\subsubsection{$e^-e^-e^+$ and $\mu^+\mu^+\mu^-$}\label{subsubsec:eee}

\begin{table}[htbp]
\renewcommand{\arraystretch}{1.15}
\centering
\scriptsize
\setlength{\tabcolsep}{2pt}
\caption{\label{tab:eee} Complex energies $E-i\Gamma/2$ of the $e^-e^-e^+$
bound and resonant states below the second threshold
$\mathrm{Ps}(2)+e^-$, grouped by the quantum numbers ($S_{12}$, $L$) and labeled
by the term symbol $^{2S_{12}+1}L^{\pi}$. Energies are in eV and half-widths
$\Gamma/2$ in meV. ``B''/``R'' denote bound/resonant states.
In the two rightmost columns, results from the literature are
presented for comparison, obtained with explicitly correlated
Gaussians~\cite{suzuki2004stochastic,matyus2013resonances}, Hylleraas
functions~\cite{bhatia1990pwave}, and correlated
exponentials~\cite{frolov1999psminus,kar2019calculations,kar2020triplet}.}
\begin{ruledtabular}
\begin{tabular}{lcclclc}
 & & & \multicolumn{2}{c}{This work} & \multicolumn{2}{c}{Lit.} \\
 & $v$ & type & \multicolumn{1}{c}{$E$} & $\Gamma/2$ & \multicolumn{1}{c}{$E$} & $\Gamma/2$ \\
\hline
\multicolumn{7}{l}{$^1S^e\ (0,0)$}\\
 & 0 & B & $-7.1295$ & $0.00$      & $-7.1295$~\cite{frolov1999psminus} & $0.00$ \\[2pt]
$\bm{\mathrm{Ps}(1)+e^-}$ & & & $\bm{-6.8028}$ & & & \\[2pt]
 & 0 & R & $-2.0689$ & $0.58$  & $-2.0689$~\cite{suzuki2004stochastic} & $0.59$ \\
 & 1 & R & $-1.7320$ & $0.12$  & $-1.7319$~\cite{suzuki2004stochastic} & $0.13$ \\
 & 2 & R & $-1.7032$ & $0.15$ & $-1.7037$~\cite{matyus2013resonances} & $0.68$ \\
\hline
\multicolumn{7}{l}{$^3S^e\ (1,0)$}\\
 & 0 & R & $-1.7289$ & $0.00$  & $-1.7289$~\cite{suzuki2004stochastic} & $0.00$ \\
 & 1 & R & $-1.7027$ & $0.26$ & $-1.7032$~\cite{matyus2013resonances} & $0.01$ \\
\hline
\multicolumn{7}{l}{$^1P^o\ (0,1)$}\\
 & 0 & R & $-1.7186$ & $0.01$  & $-1.7186$~\cite{kar2019calculations,bhatia1990pwave} & $0.01$ \\
 & 1 & R & $-1.7015$ & $0.10$  & $-1.7019$~\cite{kar2019calculations} & $0.00$ \\
\hline
\multicolumn{7}{l}{$^3P^o\ (1,1)$}\\
 & 0 & R & $-1.9954$ & $1.77$  & $-1.9953$~\cite{kar2020triplet,bhatia1990pwave} & $1.74$ \\
 & 1 & R & $-1.7170$ & $0.23$  & $-1.7170$~\cite{kar2020triplet,bhatia1990pwave} & $0.22$ \\
 & 2 & R & $-1.7014$ & $0.10$  & $-1.7018$~\cite{kar2020triplet} & $0.08$ \\[2pt]
$\bm{\mathrm{Ps}(2)+e^-}$ & & & $\bm{-1.7007}$ & & & \\[2pt]
\end{tabular}
\end{ruledtabular}
\end{table}

The negative positronium ion $\mathrm{Ps}^-$ is the lightest
three-lepton bound system. Its relevant dissociation channels are
$\mathrm{Ps}(n)+e^-$, with threshold energies given by the corresponding
positronium energy levels in Table~\ref{tab:twobody}.
The complex energies of all
states below the second threshold $\mathrm{Ps}(2)+e^-$ are collected in
Table~\ref{tab:eee} and shown in Fig.~\ref{fig:eee}. In the tables below, the
column $v$ labels the state index within a fixed spin--orbital channel. Where the
Born--Oppenheimer description applies, the same index $v$ also enumerates the
vibrational levels.

In the $^1S^e$ channel ($S_{12}=0,\,L=0$), the only bound state of
$\mathrm{Ps}^-$ appears at $E=-7.1295~\mathrm{eV}$. It lies
$0.327~\mathrm{eV}$ below the $\mathrm{Ps}(1)+e^-$ threshold and agrees with the
correlated-exponential calculation~\cite{frolov1999psminus}. The
remaining states below $\mathrm{Ps}(2)+e^-$ are resonances. Their positions
agree with previous calculations using explicitly correlated
Gaussians~\cite{suzuki2004stochastic,matyus2013resonances},
Hylleraas functions~\cite{bhatia1990pwave},
and correlated exponentials~\cite{kar2019calculations,kar2020triplet}.

\begin{figure*}[t]
\centering
\includegraphics[width=\textwidth]{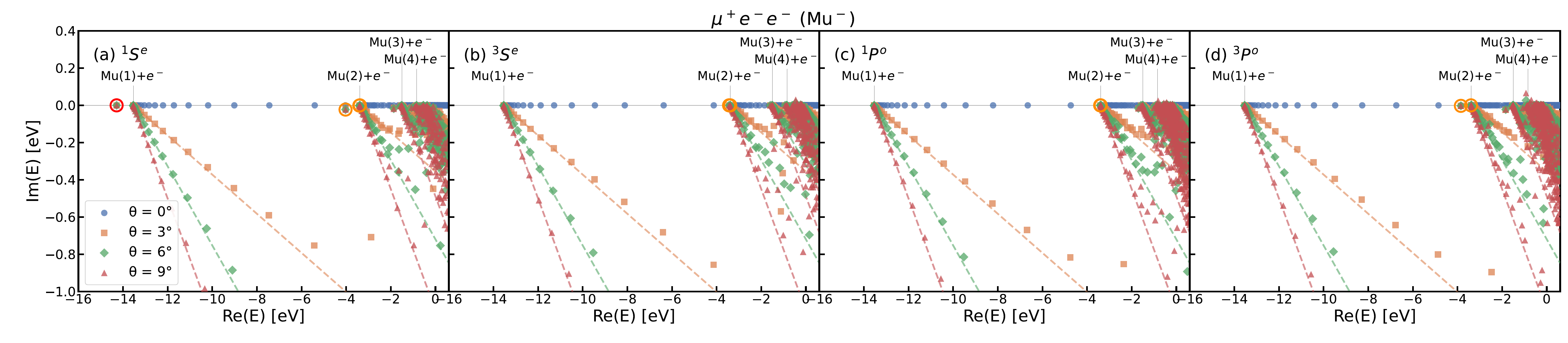}
\caption{\label{fig:EEm} Complex-energy spectra of the $\mu^+e^-e^-$
system under complex scaling for the four natural-parity channels
(a)~$^1S^e$, (b)~$^3S^e$, (c)~$^1P^o$, and (d)~$^3P^o$. Continuum eigenvalues
align along $\mathrm{Arg}(E)=-2\theta$. Stable bound and resonant poles are
circled in red and orange, respectively.}
\end{figure*}

The widths reported here are the rearrangement (fall-apart) widths for decay into
$\mathrm{Ps}(1)+e^-$, of order $10^{-2}~\mathrm{eV}$. The $e^+e^-$ annihilation
contribution ($\Gamma_{2\gamma}\sim10^{-6}~\mathrm{eV}$~\cite{Karshenboim:2003vs,Frolov:2009qi})
is several orders of magnitude below the rearrangement widths reported here.

The muonic counterpart $\mu^+\mu^+\mu^-$ is identical to $e^-e^-e^+$ up to the
overall scale $m_\mu/m_e$: as the Coulomb Hamiltonian carries no dimensionful
parameter other than the constituent mass, all energies and widths scale by
$m_\mu/m_e$ and all lengths by its inverse. Its spectrum, obtained by rescaling the spectrum in
Table~\ref{tab:eee}, is listed in Table~\ref{tab:mmm}
(Appendix~\ref{app:mmm}).

%\subsubsection{$\mu^+e^-e^-$}\label{subsubsec:Mum}

\subsection{$\mu^+e^-e^-$}\label{subsubsec:Mum}

\begin{table}[htbp]
\renewcommand{\arraystretch}{1.15}
\centering
\scriptsize
\setlength{\tabcolsep}{2pt}
\caption{\label{tab:EEm} Complex energies $E-i\Gamma/2$ of the $\mu^+e^-e^-$
bound and resonant states below the second threshold $\mathrm{Mu}(2)+e^-$,
grouped by the quantum numbers ($S_{12}$, $L$) and labeled by the term symbol
$^{2S_{12}+1}L^{\pi}$. Energies are in eV and half-widths $\Gamma/2$ in meV.
``B''/``R'' denote bound/resonant states. An asterisk ($^{*}$) marks widths
that are not fully $\theta$-converged and are only qualitative estimates. In the two rightmost columns, results
from the literature are presented for comparison, obtained with the
correlated-exponential method~\cite{liverts2013three,Frolov:2017tvw}.}
\begin{ruledtabular}
\begin{tabular}{lcclclc}
 & & & \multicolumn{2}{c}{This work} & \multicolumn{2}{c}{Lit.} \\
 & $v$ & type & \multicolumn{1}{c}{$E$} & $\Gamma/2$ & \multicolumn{1}{c}{$E$} & $\Gamma/2$ \\
\hline
\multicolumn{7}{l}{$^1S^e\ (0,0)$}\\
 & 0 & B & $-14.2875$ & $0.00$      & $-14.2875$~\cite{liverts2013three,Frolov:2017tvw} & $0.00$ \\[2pt]
$\bm{\mathrm{Mu}(1)+e^-}$ & & & $\bm{-13.5402}$ & & & \\[2pt]
 & 0 & R & $-4.0291$ & $22.70$ & $-4.0288$~\cite{liverts2013three} & $23.28$ \\
 & 1 & R & $-3.4127$ & $1.25$  & \multicolumn{1}{c}{---} & --- \\
 & 2 & R & $-3.3863$ & $0.07^{*}$ & \multicolumn{1}{c}{---} & --- \\
\hline
\multicolumn{7}{l}{$^3S^e\ (1,0)$}\\
 & 0 & R & $-3.4419$ & $0.01$  & \multicolumn{1}{c}{---} & --- \\
 & 1 & R & $-3.3881$ & $0.07^{*}$ & \multicolumn{1}{c}{---} & --- \\
\hline
\multicolumn{7}{l}{$^1P^o\ (0,1)$}\\
 & 0 & R & $-3.4134$ & $0.02$  & \multicolumn{1}{c}{---} & --- \\
 & 1 & R & $-3.3856$ & $0.11^{*}$ & \multicolumn{1}{c}{---} & --- \\
\hline
\multicolumn{7}{l}{$^3P^o\ (1,1)$}\\
 & 0 & R & $-3.8493$ & $2.68$  & \multicolumn{1}{c}{---} & --- \\
 & 1 & R & $-3.3962$ & $0.06$  & \multicolumn{1}{c}{---} & --- \\[2pt]
$\bm{\mathrm{Mu}(2)+e^-}$ & & & $\bm{-3.3851}$ & & & \\[2pt]
\end{tabular}
\end{ruledtabular}
\end{table}

The negative muonium ion $\mathrm{Mu}^-$ is the muonic counterpart of
$\mathrm{Ps}^-$. It consists of two identical electrons bound to an antimuon.
The dissociation channels are $\mathrm{Mu}(n)+e^-$. The corresponding threshold
energies are given by the muonium energy levels in Table~\ref{tab:twobody},
approximately twice the corresponding $\mathrm{Ps}(n)$ threshold energies. The bound and resonant
states
below $\mathrm{Mu}(2)+e^-$ are listed in Table~\ref{tab:EEm} and shown in
Fig.~\ref{fig:EEm}.

The $^1S^e$ channel contains one bound state at
$E=-14.2875~\mathrm{eV}$. It lies $0.747~\mathrm{eV}$ below
$\mathrm{Mu}(1)+e^-$ threshold. The obtained binding energy agrees with the correlated-exponential
calculations~\cite{liverts2013three,Frolov:2017tvw}. The energy of lowest $^1S^e$ resonance
also consistent with the result of Ref.~\cite{liverts2013three}.
We also resolve two shallow states just below $\mathrm{Mu}(2)+e^-$, one in each spin channel. To our knowledge, their complex energies have not been reported previously.
Each $P$-wave channel also contains a shallow resonance
near the $\mathrm{Mu}(2)+e^-$ threshold.

Since $m_\mu/m_e\simeq207$, the antimuon is nearly static on the electronic motion scale.
Thus $\mathrm{Mu}^-$ is close to a two-electron atom with a heavy positive
center, analogous to $\mathrm{H}^-$ atomic configuration. The strongest attraction occurs in the
$^1S^e$ channel. This channel has no centrifugal barrier, and the electron-spin
singlet requires a symmetric spatial wave function. Both features enhance the
electron localization around the positive center.
%The symmetric spatial state also gives stronger $e^-e^-$ Coulomb repulsion than the
%triplet channel. However, dynamical correlation forms a Coulomb hole and partly
%lowers the average repulsion.
Among the lowest ($v=0$) resonances below $\mathrm{Mu}(2)$, the $^1S^e$ and
$^3P^o$ poles lie within about $0.2~\mathrm{eV}$ of each other.
However the half-width of $^1S^e$
resonance is roughly an order of magnitude larger than that of $^3P^o$ resonance. The $^3S^e$ and
$^1P^o$ resonances form the next near-degenerate pair and have comparable widths.
From lowest to highest energy, the $v=0$ resonances follow the same channel ordering as in $\mathrm{Ps}^-$,
$E(^1S^e)\lesssim E(^3P^o)<E(^3S^e)\lesssim E(^1P^o)$.

%\subsubsection{$\mu^+\mu^+e^-$}\label{subsubsec:mmE}

\subsection{$\mu^+\mu^+e^-$}\label{subsubsec:mmE}

\begin{figure*}[t]
\centering
\includegraphics[width=\textwidth]{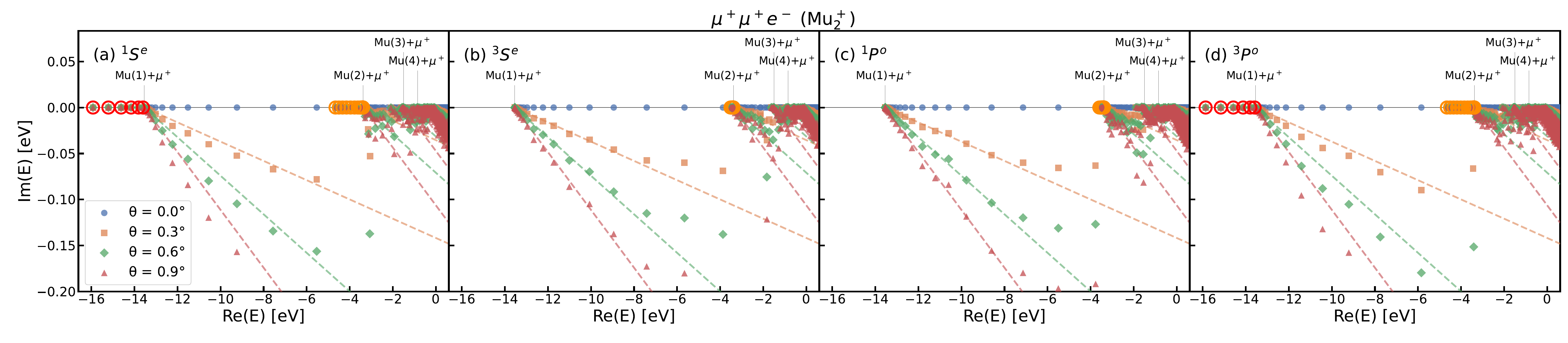}
\caption{\label{fig:mmE} Complex-energy spectra of the $\mu^+\mu^+e^-$
system under complex scaling for the four natural-parity channels
(a)~$^1S^e$, (b)~$^3S^e$, (c)~$^1P^o$, and (d)~$^3P^o$. Continuum eigenvalues
align along $\mathrm{Arg}(E)=-2\theta$. Stable bound and resonant poles are
circled in red and orange, respectively.}
\end{figure*}

Unlike the atomic $\mathrm{Ps}^-$ and $\mathrm{Mu}^-$ systems,
$\mu^+\mu^+e^-$ is conventionally regarded as a molecular ion. It contains two heavy $\mu^+$
centers and one light electron, making it the muonic analogue of
$\mathrm{H}_2^+$.
Its dissociation channels are $\mathrm{Mu}(n)+\mu^+$. The corresponding threshold
energies are given by the same muonium level energies as for
$\mathrm{Mu}^-$ (Table~\ref{tab:twobody}). The bound and resonant states below
$\mathrm{Mu}(2)+\mu^+$ are listed in Table~\ref{tab:mmE} and shown in
Fig.~\ref{fig:mmE}.

The Born--Oppenheimer picture describes the bound states spectrum of $\mu^+\mu^+e^-$
below the lowest dissociation threshold. In this picture, the electron moves in the field of two
slowly moving $\mu^+$ centers. The electronic eigenenergies as functions of the $\mu^+\mu^+$
separation define Born--Oppenheimer potential curves for the relative motion of the two muons.
We label these electronic curves by the usual
one-electron molecular notation ${}^{2S_e+1}\Sigma_{g/u}^{+}$, with
$S_e=1/2$. This notation labels the BO electronic curve, whereas the table
labels $^{2S_{\mu\mu}+1}L^\pi$ refer to the exact quantum numbers. The
$^2\Sigma_g^+$ electronic curve approaches the
$\mathrm{Mu}(1)+\mu^+$ dissociation limit asymptotically and supports bound vibrational levels. Because the two
$\mu^+$ are identical fermions, the muon-pair spin $S_{\mu\mu}$ is tied to the
parity of the $\mu^+$--$\mu^+$ rotational wave function. For the $^2\Sigma_g^+$
configuration, the
even- and odd-$L$ correspond to the $^1S^e$ and $^3P^o$ channels. The large reduced mass of the muon pair suppresses the rotational splittings.
Consequently, the bound states supported by the $^2\Sigma_g^+$ curve appear in both $^1S^e$ and $^3P^o$ channels. The antibonding $^2\Sigma_u^+$ curve is overall repulsive but
has a shallow long-range minimum. In the hydrogen
molecular ion $\mathrm{H}_2^+$ this minimum supports one weakly bound vibrational
level~\cite{hilico2000ab}. However, the muon is much lighter than the proton, so that the
vibrational zero-point energy of the $\mu^+$--$\mu^+$ motion exceeds the depth
of this minimum. We therefore find no bound $^3S^e$ or $^1P^o$ state.

The $^2\Sigma_g^+$ curve supports six vibrational levels in both the $^1S^e$ and
$^3P^o$ channels. These BO paired bound states extended from the $^1S^e$ ground state at
$E=-15.9221~\mathrm{eV}$ ($v=0$) to the shallow $v=5$ states just below
$\mathrm{Mu}(1)+\mu^+$ threshold. The Feshbach resonances below the
$\mathrm{Mu}(2)+\mu^+$ threshold belong to the excited electronic channels.
For the lowest ($v=0$) such resonances, the channel ordering differs slightly
from that in the two atomic systems, $\mathrm{Ps}^-$ and $\mathrm{Mu}^-$,
with the $^3S^e$ and $^1P^o$ channels interchanged,
$E(^{1}S^e)\lesssim E(^{3}P^o)<E(^{1}P^o)\lesssim E(^{3}S^e).$

The resonance half-widths are difficult to determine robustly in this system. The reason is as follows. The
internal reduced mass of the $\mathrm{Mu}$ is of order
$m_e$. But the external reduced mass of the muon spectator relative to the muonium
atom is of order $m_\mu/2$. These scales differ by about two orders of magnitude.
Stabilizing the
complex-scaled pole positions therefore requires basis functions that describe both
compact electronic motion and extended muon--muon motion. This makes the pole
extraction sensitive to the basis-set range. We therefore use small rotation
angles, $\theta=0.3,0.6,0.9^\circ$, for the resonance search. The real parts are
stable under these rotations. The imaginary parts are not, and sub-meV
half-widths cannot be resolved reliably. We therefore list only the real part of
the complex energy in Table~\ref{tab:mmE}.

\begin{table*}[t]
\renewcommand{\arraystretch}{1.05}
\setlength{\tabcolsep}{4pt}
\centering
\footnotesize
\caption{\label{tab:mmE} Complex energies of the $\mu^+\mu^+e^-$
bound and resonant states below the second threshold $\mathrm{Mu}(2)+\mu^+$,
grouped by the quantum numbers ($S_{\mu\mu}$, $L$) and labeled by the term symbol
$^{2S_{\mu\mu}+1}L^{\pi}$. Energies $E$ are in eV, and $v$ labels the state
index within each channel. ``B''/``R'' denote bound/resonant states. In the rightmost
column, results from the literature for the $^1S^e$ bound states are presented for
comparison, obtained with the correlated-exponential
method~\cite{liverts2013three}. For the resonances, only the real part of the
complex energy is listed because the small rotation angles do not resolve the
sub-meV half-widths reliably.}
\begin{tabular*}{0.49\textwidth}[t]{@{\extracolsep{\fill}}lrccc}
\hline\hline & & & \multicolumn{1}{c}{This work} & \multicolumn{1}{c}{Lit.} \\
 & $v$ & type & \multicolumn{1}{c}{$E$} & \multicolumn{1}{c}{$E$} \\\hline
\multicolumn{5}{l}{$^1S^e\ (0,0)$}\\
 & 0 & B & $-15.9221$ & $-15.9221$~\cite{liverts2013three} \\
 & 1 & B & $-15.2065$ & $-15.2065$~\cite{liverts2013three} \\
 & 2 & B & $-14.6186$ & $-14.6187$~\cite{liverts2013three} \\
 & 3 & B & $-14.1528$ & --- \\
 & 4 & B & $-13.8104$ & --- \\
 & 5 & B & $-13.6011$ & --- \\
$\bm{\mathrm{Mu}(1){+}\mu^+}$ & & & $\bm{-13.5402}$ & \\
 & 0 & R & $-4.6609$ & --- \\
 & 1 & R & $-4.5033$ & --- \\
 & 2 & R & $-4.3344$ & --- \\
 & 3 & R & $-4.1540$ & --- \\
 & 4 & R & $-3.9660$ & --- \\
 & 5 & R & $-3.7767$ & --- \\
 & 6 & R & $-3.6243$ & --- \\
 & 7 & R & $-3.5251$ & --- \\
 & 8 & R & $-3.4687$ & --- \\
 & 9 & R & $-3.4371$ & --- \\
 & 10 & R & $-3.4177$ & --- \\
 & 11 & R & $-3.4053$ & --- \\
 & 12 & R & $-3.3973$ & --- \\
 & 13 & R & $-3.3924$ & --- \\
 & 14 & R & $-3.3893$ & --- \\
 & 15 & R & $-3.3874$ & --- \\
 & 16 & R & $-3.3863$ & --- \\
 & 17 & R & $-3.3856$ & --- \\
 & 18 & R & $-3.3852$ & --- \\
\hline
\multicolumn{5}{l}{$^3S^e\ (1,0)$}\\
 & 0 & R & $-3.5223$ & --- \\
 & 1 & R & $-3.4920$ & --- \\
 & 2 & R & $-3.4635$ & --- \\
 & 3 & R & $-3.4395$ & --- \\
 & 4 & R & $-3.4210$ & --- \\
 & 5 & R & $-3.4077$ & --- \\
 & 6 & R & $-3.3990$ & --- \\
 & 7 & R & $-3.3936$ & --- \\
 & 8 & R & $-3.3900$ & --- \\
 & 9 & R & $-3.3877$ & --- \\
 & 10 & R & $-3.3863$ & --- \\
 & 11 & R & $-3.3856$ & --- \\
$\bm{\mathrm{Mu}(2){+}\mu^+}$ & & & $\bm{-3.3851}$ & \\
\hline\hline\end{tabular*}
\hfill
\begin{tabular*}{0.49\textwidth}[t]{@{\extracolsep{\fill}}lrccc}
\hline\hline & & & \multicolumn{1}{c}{This work} & \multicolumn{1}{c}{Lit.} \\
 & $v$ & type & \multicolumn{1}{c}{$E$} & \multicolumn{1}{c}{$E$} \\\hline
\multicolumn{5}{l}{$^3P^o\ (1,1)$}\\
 & 0 & B & $-15.8623$ & --- \\
 & 1 & B & $-15.1560$ & --- \\
 & 2 & B & $-14.5768$ & --- \\
 & 3 & B & $-14.1198$ & --- \\
 & 4 & B & $-13.7868$ & --- \\
 & 5 & B & $-13.5883$ & --- \\
$\bm{\mathrm{Mu}(1){+}\mu^+}$ & & & $\bm{-13.5402}$ & \\
 & 0 & R & $-4.6584$ & --- \\
 & 1 & R & $-4.5049$ & --- \\
 & 2 & R & $-4.3506$ & --- \\
 & 3 & R & $-4.1866$ & --- \\
 & 4 & R & $-4.0157$ & --- \\
 & 5 & R & $-3.8485$ & --- \\
 & 6 & R & $-3.6929$ & --- \\
 & 7 & R & $-3.5743$ & --- \\
 & 8 & R & $-3.4991$ & --- \\
 & 9 & R & $-3.4564$ & --- \\
 & 10 & R & $-3.4313$ & --- \\
 & 11 & R & $-3.4155$ & --- \\
 & 12 & R & $-3.4050$ & --- \\
 & 13 & R & $-3.3979$ & --- \\
 & 14 & R & $-3.3932$ & --- \\
 & 15 & R & $-3.3902$ & --- \\
 & 16 & R & $-3.3882$ & --- \\
 & 17 & R & $-3.3869$ & --- \\
 & 18 & R & $-3.3860$ & --- \\
 & 19 & R & $-3.3854$ & --- \\
\hline
\multicolumn{5}{l}{$^1P^o\ (0,1)$}\\
 & 0 & R & $-3.5956$ & --- \\
 & 1 & R & $-3.5221$ & --- \\
 & 2 & R & $-3.5123$ & --- \\
 & 3 & R & $-3.4931$ & --- \\
 & 4 & R & $-3.4673$ & --- \\
 & 5 & R & $-3.4451$ & --- \\
 & 6 & R & $-3.4444$ & --- \\
 & 7 & R & $-3.4267$ & --- \\
 & 8 & R & $-3.4133$ & --- \\
 & 9 & R & $-3.4039$ & --- \\
 & 10 & R & $-3.3974$ & --- \\
 & 11 & R & $-3.3966$ & --- \\
 & 12 & R & $-3.3930$ & --- \\
 & 13 & R & $-3.3900$ & --- \\
 & 14 & R & $-3.3881$ & --- \\
 & 15 & R & $-3.3868$ & --- \\
 & 16 & R & $-3.3858$ & --- \\
 & 17 & R & $-3.3852$ & --- \\
$\bm{\mathrm{Mu}(2){+}\mu^+}$ & & & $\bm{-3.3851}$ & \\
\hline\hline\end{tabular*}\ \null% the trailing "\ \null" reproduces the exact panel spacing of the former \input{...} layout
\end{table*}

\begin{figure*}[t]
\centering
\includegraphics[width=\textwidth]{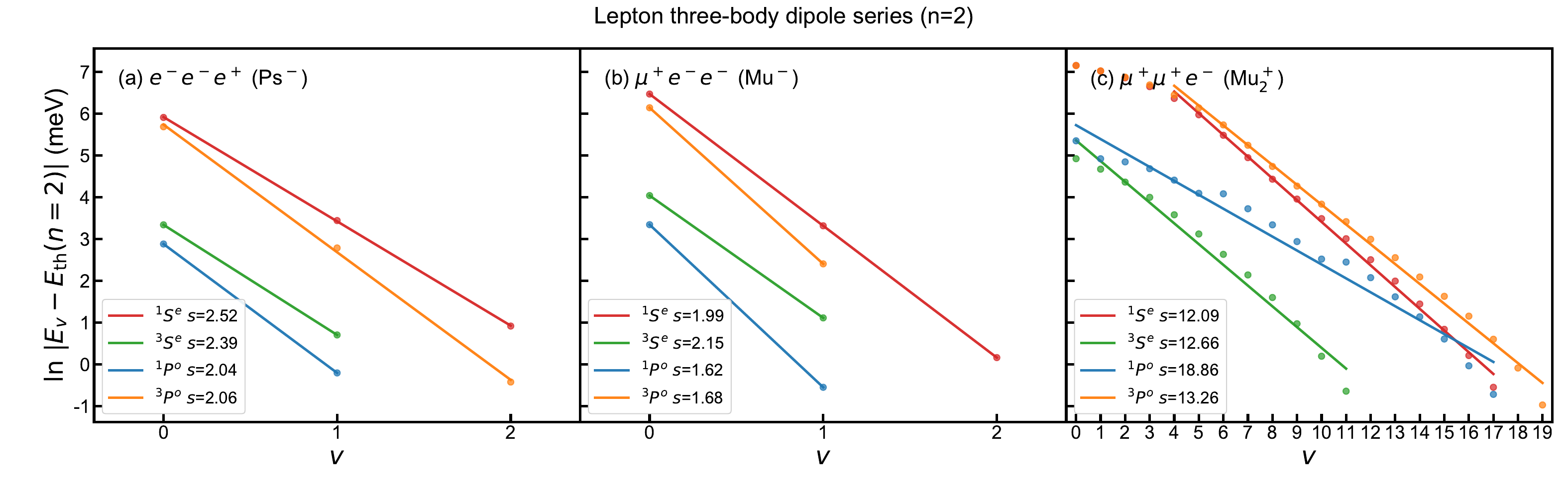}
\caption{\label{fig:GD} Gailitis--Damburg dipole series at the $n=2$ threshold
for the three trilepton systems: (a) $\mathrm{Ps}^-$, (b) $\mathrm{Mu}^-$, and
(c) $\mathrm{Mu}_2^+$. For each spin--parity channel,
$\ln|E_v-E_{\rm th}(n{=}2)|$ is plotted against the state index $v$.
A fit to each channel gives the effective dipole
parameter $s$ listed in the legend.}
\end{figure*}

\section{Universal dynamics of trilepton systems at $n=2$ thresholds}\label{sec:GD}

The three trilepton systems have distinct conventional structural descriptions.
 In the limit $m_\mu\to\infty$, $\mathrm{Mu}^-$ reduces to an atomic configuration with one fixed positive center.
 By contrast, $\mathrm{Mu}_2^+$ reduces to a molecular configuration with two fixed positive centers.
 The equal-mass $\mathrm{Ps}^-$ has no analogous separation of heavy and light constituents.
Despite these different limiting configurations, the spectra in Sec.~\ref{sec:results} contain common near threshold behavior.
All three systems contain multiple Feshbach resonances below their degenerate $n=2$
thresholds. There are especially many shallow resonances. By contrast, The bound states spectra show no accumulation below the lowest dissociation thresholds.

This common concentration of shallow resonances suggests a connection to the
Gailitis--Damburg mechanism~\cite{gailitis1963,gailitis1982finite}. The mechanism arises near a degenerate threshold.
The external lepton mixes the $2S$ and $2P$ states
of the neutral two-body atom through the linear Stark effect, producing an
attractive inverse-square charge--dipole interaction. When this attraction is
supercritical, it supports a geometrically spaced sequence of resonances that accumulates
at the threshold.

At first sight, this picture most closely resembles the
one-heavy--two-light atomic configuration. Its realization in the finite
spectra of the equal-mass and two-heavy--one-light systems is less obvious,
particularly for the conventionally molecular $\mathrm{Mu}_2^+$ system.
A systematic numerical comparison across all three mass configurations is
therefore needed.
To perform this comparison, we first derive the asymptotic dipole parameters
for the natural-parity $S$- and $P$-wave channels. We then compare the
predicted geometric scaling with the numerical spectra.

\subsection{Gailitis--Damburg mechanism}\label{subsec:GD_dipole}

Particles $1$ and $2$ form the neutral two-body atom $X$, and particle $3$ is the
spectator. The relevant reduced masses are
\begin{equation}
\mu_X=\frac{m_1m_2}{m_1+m_2},\qquad
\mu_R=\frac{m_3(m_1+m_2)}{m_1+m_2+m_3}.
\end{equation}
The Bohr length of atom $X$ is
\begin{equation}
a_X=(\alpha\mu_X)^{-1}.
\end{equation}
At the degenerate $n=2$ threshold, the leading long-range interaction is the
charge--dipole interaction
\begin{equation}\label{eq:GDVdip}
V_{\rm dip}=-\alpha Q_3\,\frac{\bm r\!\cdot\!\bm R}{R^3},
\end{equation}
where $\bm r$ and $\bm R$ are the internal and atom--spectator coordinates, and
$Q_3$ is the spectator charge. In the basis $|(\ell,\lambda)LM\rangle$
($\ell=0,1$), the large-$R$ radial equation is
\begin{equation}\label{eq:GDschrmat}
\begin{aligned}
\sum_j \bigg[
&-\frac{\delta_{ij}}{2\mu_R}\frac{d^2}{dR^2}
+\frac{\Lambda_{ij}}{2\mu_R R^2}
-\frac{D_{ij}}{R^2}
\bigg]u_j(R) = E\,u_i(R),
\end{aligned}
\end{equation}
where $i=(\ell_i,\lambda_i)$ and
\begin{equation}
\Lambda_{ij}=\lambda_i(\lambda_i+1)\delta_{ij}.
\end{equation}
The dipole matrix is
\begin{equation}\label{eq:GDD}
D_{ij}
=\alpha Q_3\,\langle 2\ell_i|r|2\ell_j\rangle A_{ij},
\end{equation}
with the angular factor $A_{ij}$ given in Appendix~\ref{app:GD_angular}. The
nonzero radial coupling is
\begin{equation}\label{eq:GDradial}
\langle 2S|r|2P\rangle
=\int_0^\infty R_{20}(r)\,r\,R_{21}(r)\,r^2\,dr
=-3\sqrt3\,a_X.
\end{equation}
After multiplying Eq.~\eqref{eq:GDschrmat} by $-2\mu_R$, one diagonalizes
$M_{ij}=-\Lambda_{ij}+2\mu_R D_{ij}$. An eigenvalue $x_k>1/4$, corresponding
to a supercritical inverse-square channel, gives
\begin{equation}\label{eq:GDs}
s_k\equiv\sqrt{x_k-\frac14}.
\end{equation}
For such channels, scale invariance makes the near-threshold spectrum
geometric. The most attractive channel,
$s=\max_k s_k$, gives
\begin{equation}\label{eq:GD_asymptotic_ratio}
\frac{|E_{v+1}-E_{\rm th}|}{|E_{v}-E_{\rm th}|}=e^{-2\pi/s}.
\end{equation}
For $L=0$, the channels $(\ell,\lambda)=(0,0)$ and $(1,1)$ give, with
$\eta=\mu_R/\mu_X$,
\begin{equation}\label{eq:GDL0}
M=
\begin{pmatrix}
0 & 6\eta\\
6\eta & -2
\end{pmatrix}.
\end{equation}
Thus
\begin{equation}\label{eq:GDL0s}
x_{\max}=-1+\sqrt{1+36\eta^2},\qquad
s=\sqrt{-\frac54+\sqrt{1+36\eta^2}}.
\end{equation}
For $L=1$, the coupled block $(\ell,\lambda)=(0,1),(1,0),(1,2)$ gives
\begin{equation}\label{eq:GDL1s}
x_{\max}=-3+3\sqrt{1+4\eta^2},\qquad
s=\sqrt{-\frac{13}{4}+3\sqrt{1+4\eta^2}}.
\end{equation}

\subsection{Realization of the Gailitis--Damburg mechanism}\label{example:GD}

We now apply the above analysis to the resolved resonance spectra below the second
threshold, $X(2)+\ell$, in Figs.~\ref{fig:eee}, \ref{fig:EEm}, and
\ref{fig:mmE}. The analytic dipole-limit values of $s$ are obtained from
Eqs.~\eqref{eq:GDL0s} and~\eqref{eq:GDL1s}. Since these values depend on
$\eta=\mu_R/\mu_X$ rather than on the identical-pair spin, Table~\ref{tab:GD}
compares one dipole-limit prediction for each $L$ with the separate singlet and
triplet fits.

\begin{table}[htbp]
\centering
\caption{\label{tab:GD} Dipole parameter $s$ at the $n=2$ threshold. The table
compares the analytic dipole-limit values with the singlet and triplet fits for
the $S$- and $P$-wave channels.}
\begin{ruledtabular}
\begin{tabular}{lccc}
\multirow{2}{*}{system} & \multirow{2}{*}{$s$ (dipole limit)} & \multicolumn{2}{c}{$s$ (fit)} \\
 &  & singlet & triplet \\
\hline
$\mathrm{Mu}^-\ (L{=}0)$   & $2.20$  & $1.99$  & $2.15$  \\
$\mathrm{Mu}^-\ (L{=}1)$   & $1.86$  & $1.62$  & $1.68$  \\
$\mathrm{Ps}^-\ (L{=}0)$   & $2.61$  & $2.52$  & $2.39$  \\
$\mathrm{Ps}^-\ (L{=}1)$   & $2.30$  & $2.04$  & $2.06$  \\
$\mathrm{Mu}_2^+\ (L{=}0)$ & $24.97$ & $12.09$ & $12.66$ \\
$\mathrm{Mu}_2^+\ (L{=}1)$ & $24.93$ & $18.86$ & $13.26$ \\
\end{tabular}
\end{ruledtabular}
\end{table}

Fig.~\ref{fig:GD} characterizes the resolved sequences by plotting
$\ln|E_v-E_{\mathrm{th}}|$ against $v$. Within each channel, the resolved
states follow an approximately linear trend. Fitting each slope to
$-2\pi/s_{\mathrm{fit}}$ yields an effective scaling parameter
$s_{\mathrm{fit}}$.

In $\mathrm{Ps}^-$ and $\mathrm{Mu}^-$, the electron spectator gives
$\eta=4/3$ and $\eta\simeq1$, respectively. For $L=0$, the corresponding
dipole-limit ratios are $e^{-2\pi/s}\simeq0.09$ and $0.06$. The $L=1$
ratios are even smaller. Thus, the detuning from threshold decreases by
more than an order of magnitude between successive resonances. Higher
members rapidly approach the threshold and become difficult to resolve
numerically.

In $\mathrm{Mu}_2^+$, the heavy $\mu^+$ spectator gives
$\eta\simeq104$. The analytic dipole-limit values are $s\simeq25$ for
both $L=0$ and $L=1$, corresponding to a near-unit geometric ratio
$e^{-2\pi/s}\simeq0.78$. This produces dense sequences immediately below
the $\mathrm{Mu}(2)+\mu^+$ threshold. We resolve $19$ members in the
$^1S^e$ channel, $20$ in the $^3P^o$ channel, and additional sequences
in the $^3S^e$ and $^1P^o$ channels (Table~\ref{tab:mmE}).

The resolved $\mathrm{Mu}_2^+$ states yield the smaller values
$s_{\mathrm{fit}}\simeq12$--$19$. The large atom--spectator reduced mass
compresses the radial scale at a given detuning relative to the
electron-spectator systems. These states therefore probe smaller
atom--spectator separations, where the dipole attraction is present but
shorter-range effects still modify the level spacings. The fitted values
should thus be interpreted as finite-$R$ effective parameters rather
than asymptotic constants. The $P$-wave fits lie closer to the dipole
limit than the $S$-wave fits, particularly in the singlet sector. The centrifugal barrier
may explain this trend by reducing the sensitivity of the $P$-wave states to short-range dynamics.

Fig.~\ref{fig:GD} also reveals a crossover in the spectral organization.
At small $v$, corresponding to Feshbach resonances lying deeper below the
$n=2$ threshold, the $^1S^e$ and $^3P^o$ sequences lie close in energy.
A similar pairing occurs between the $^3S^e$ and $^1P^o$ channels.
This pattern is particularly evident in $\mathrm{Mu}_2^+$, where the
first four resonances of the $^1S^e$ and $^3P^o$ channels nearly coincide
pairwise. These low $v$ pairings are consistent with the
Born--Oppenheimer organization of the deeper resonance spectrum.
As $v$ increases, the resonances approach the $n=2$ threshold and the
Gailitis--Damburg scaling becomes dominant. The $\mathrm{Mu}_2^+$ spectrum
therefore cannot be understood solely from a molecular viewpoint. The
Born--Oppenheimer picture organizes the deeper spectrum, whereas the
near-threshold resonances are governed by the atomic
$\mathrm{Mu}(2)+\mu^+$ channel and its charge--dipole interaction.

The Gailitis--Damburg mechanism does not work at the $n=1$ thresholds.
The ground-state two-body atom has no degenerate opposite-parity partner,
so the first-order dipole coupling is absent. The leading attraction is
instead a shorter-ranged polarization potential proportional to
$-1/R^4$, which does not support a geometric series.

\section{Tetralepton systems}\label{subsec:4l}

\subsection{$e^+e^+e^-e^-$ and $\mu^+\mu^+\mu^-\mu^-$}\label{subsubsec:eeEE}

\begin{figure*}[t]
\centering
\includegraphics[width=\textwidth]{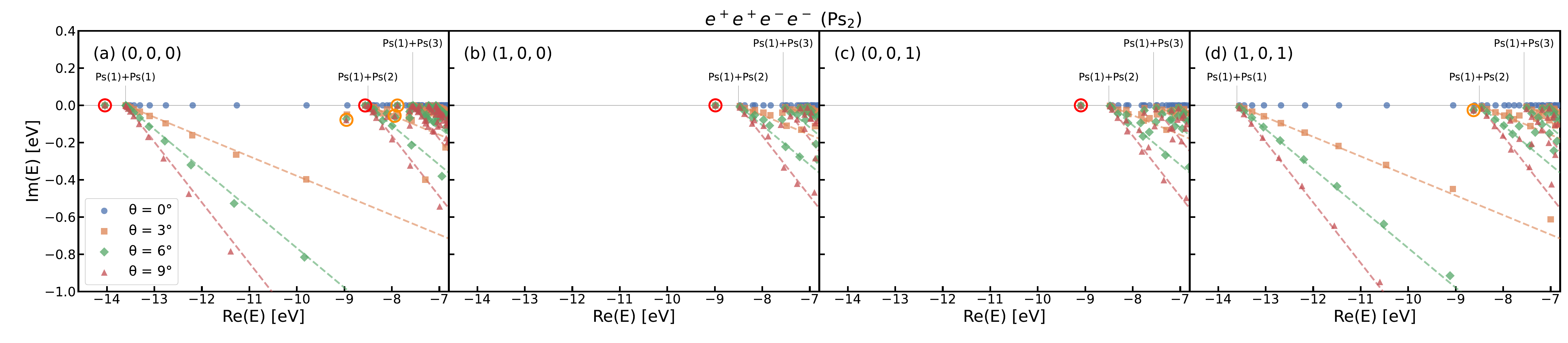}
\caption{\label{fig:eeEE} Complex-energy spectra of the $e^+e^+e^-e^-$
system under complex scaling for $\theta=0,3,6,9^\circ$, in the four spin--orbital
channels (a)~$(0,0,0)$, (b)~$(1,0,0)$, (c)~$(0,0,1)$, and (d)~$(1,0,1)$, labeled
by $(S_{12},S_{34},L)$. Continuum eigenvalues rotate along the rays
$\mathrm{Arg}(E-E_{\rm th})=-2\theta$ from each open threshold. Bound states (red
circles) and resonances (orange circles) remain fixed. The
$\mathrm{Ps}(1)+\mathrm{Ps}(1)$ threshold appears only in Pauli-allowed
channels.}
\end{figure*}

The positronium molecule $e^+e^+e^-e^-$ is an equal-mass
four-lepton system. We label each channel by
$(S_{12},S_{34},L)$, where $S_{12}$ and $S_{34}$ are the spins of the $e^+e^+$
and $e^-e^-$ pairs. Since the system has charge conjugate symmetry, $C$
interchanges the two pair spins, $(S_{12},S_{34},L)\leftrightarrow(S_{34},S_{12},L)$.
The channels $(0,1,L)$ and $(1,0,L)$ are related by charge conjugation and therefore have identical spectra.
We list only $(1,0,L)$. The remaining channels $(0,0,L)$ and $(1,1,L)$ are self-conjugate. In $\mathrm{Ps_2}$, the
dissociation channels are $\mathrm{Ps}(n)+\mathrm{Ps}(n')$, with threshold
energies given by the sums of the corresponding positronium energy levels in
Table~\ref{tab:twobody}. The
$\mathrm{Ps}(1)+\mathrm{Ps}(1)$ threshold contains two identical
$\mathrm{Ps}(1)$ atoms. Exchanging the two atoms changes the relative orbital
motion by $(-1)^L$ and the pair-spin part by $(-1)^{S_{12}+S_{34}}$. This operation simultaneously exchanges
the two electrons and the two positrons. The two fermionic exchanges together require the total wave function
to be symmetric under the exchange of the two atoms. The threshold is therefore open only when $(-1)^{S_{12}+S_{34}+L}=1.$
We refer to the first threshold as Pauli-allowed when this condition
is satisfied and as Pauli-forbidden otherwise.

The $v=0$ ground state in the $(0,0,0)$ channel is the leptonic analogue of the hydrogen molecule $\mathrm{H}_2$. We obtain
$E=-14.0410~\mathrm{eV}$, corresponding to a binding energy of
$0.435~\mathrm{eV}$ relative to the $\mathrm{Ps}(1)+\mathrm{Ps}(1)$ threshold.
This value agrees with earlier correlated-Gaussian calculations~\cite{Kinghorn:1993zz,Varga:1998ss},
$E=-14.0412~\mathrm{eV}$. When the lowest threshold is exchange
symmetry forbidden,
two additional bound states appear below $\mathrm{Ps}(1)+\mathrm{Ps}(2)$. They
are the $v=0$ state in $(0,0,1)$ at $-9.0916~\mathrm{eV}$ and the $v=0$ state in
$(1,0,0)$ at $-8.9862~\mathrm{eV}$. The $(0,0,1)$ state is the first $L=1$
excited state accessed experimentally through the electric-dipole transition
from the $\mathrm{Ps}_2$ ground state near $251~\mathrm{nm}$
($4.94~\mathrm{eV}$)~\cite{Cassidy_2012}. From the present energies, the
transition energy is $4.9494~\mathrm{eV}$. This observation provides a direct
experimental evidence for the natural-parity $P$-wave sector considered here.
The $(1,0,0)$ channel also contains a bound state at
$E=-8.9862~\mathrm{eV}$. Its energy agrees with
previous calculations that included orbital excitations
~\cite{suzuki2000excited,matyus2012molecular}, but the state was not
resolved in our previous GEM calculation with an all $S$-wave basis
~\cite{Ma:2025rvj}. To clarify the role of the omitted configurations,
we performed additional GEM calculations with selected configuration of Gaussian basis. Neither
the all $S$-wave basis nor the basis describing
$\mathrm{Ps}(1S)+\mathrm{Ps}(2P)$ with $P$-wave relative motion
reproduced the bound state alone. The state appeared only when the two
configurations are coupled. In the latter, the internal and intercluster
$P$ waves couple to total $L=0$. Because $\mathrm{Ps}(2S)$ and
$\mathrm{Ps}(2P)$ are degenerate, the two configurations correspond to the same threshold energy.
This configuration dependece reflects that both threshold are important for the formation of near-threshold
bound states and resonances.

The $(0,0,0)$ channel also contains a candidate narrow resonance just
below the $\mathrm{Ps}(1)+\mathrm{Ps}(2)$ threshold.
It lies at
$E=-8.5613~\mathrm{eV}$,
with an estimated half-width of only $0.55~\mathrm{meV}$. Its real energy agrees with
the charge-conjugation $C=-1$ bound state at $-8.5628~\mathrm{eV}$ reported
in Refs.~\cite{suzuki2004stochastic,matyus2013resonances}.
The internal excitation of the $(e^+e^+)$ and $(e^-e^-)$ must be different.
Therefore, one cannot simply calculate the C-parity as $C=(-1)^{L+S}=+1$.
The same mechanism in the $C=-1$ fully-charm tetraquark system was discussed
in detail in Appendix B of Ref.~\cite{Wu:2024euj}.
This state is forbidden to decay into the $C=+1$ $S$-wave
$\mathrm{Ps}(1)+\mathrm{Ps}(1)$ channel and
therefore has zero width. We interpret the small imaginary part in the present
calculation as a numerical residual and classify the state as a $C=-1$ bound state.

Between the two lowest thresholds, the $(0,0,0)$ channel contains two resonances.
The first is a broad resonance at $-8.9560~\mathrm{eV}$, which corresponds to the known $S$-wave
$\mathrm{Ps}_2$ Feshbach resonance~\cite{ho1989resonant,suzuki2004stochastic}.
The $(1,0,1)$ channel contains a $P$-wave resonance at $-8.6223~\mathrm{eV}$,
consistent with the previously reported resonance~\cite{suzuki2004stochastic}. No stable pole is found
below $\mathrm{Ps}(1)+\mathrm{Ps}(2)$ in the $(1,1,0)$ or $(1,1,1)$ channel.
Above $\mathrm{Ps}(1)+\mathrm{Ps}(2)$, the $(0,0,0)$ channel contains two additional resonances below $\mathrm{Ps}(1)+\mathrm{Ps}(3)$, at $-7.9413$ and
$-7.8823~\mathrm{eV}$. The complete spectra are given in Table~\ref{tab:eeEE}.

The equal-mass muonic system $\mu^+\mu^+\mu^-\mu^-$ has the same dimensionless
Coulomb spectrum. Its energies and widths are obtained from the $\mathrm{Ps}_2$
values by the overall mass scaling $m_\mu/m_e$, while lengths scale by the
inverse factor. The complete spectra are given in Appendix~\ref{app:mmm}.

\begin{table}[htbp]
\renewcommand{\arraystretch}{1.15}
\centering
\scriptsize
\setlength{\tabcolsep}{2pt}
\caption{\label{tab:eeEE} Complex energies of the $e^+e^+e^-e^-$ bound
and resonant states, grouped by $(S_{12},S_{34},L)$. Energies $E$ are in eV and
half-widths $\Gamma/2$ in meV. ``B''/``R'' denote bound/resonant states.
Literature values are shown in the two rightmost columns. The $(0,0,0)$ state
marked $^{*}$ corresponds to the
reported $C=-1$ bound state of
Refs.~\cite{suzuki2004stochastic,matyus2013resonances}. Its small listed
half-width is a numerical residual.}
\begin{ruledtabular}
\begin{tabular}{lcclclc}
 & & & \multicolumn{2}{c}{This work} & \multicolumn{2}{c}{Lit.} \\
 & $v$ & type & \multicolumn{1}{c}{$E$} & $\Gamma/2$ & \multicolumn{1}{c}{$E$} & $\Gamma/2$ \\
\hline
\multicolumn{7}{l}{$(0,0,0)$}\\
 & 0 & B & $-14.0410$ & $0.00$ & $-14.0412$~\cite{Kinghorn:1993zz,Varga:1998ss} & $0.00$ \\[2pt]
$\bm{\mathrm{Ps}(1)+\mathrm{Ps}(1)}$ & & & $\bm{-13.6057}$ & & & \\[2pt]
 & 0 & R & $-8.9560$ & $77.70$ & $-8.9629$~\cite{matyus2013resonances} & $82.45$ \\
 & 1 & B$^{*}$ & $-8.5613$ & $0.55^{*}$ & $-8.5628$~\cite{suzuki2004stochastic,matyus2013resonances} & $0.00$ \\
\hline
\multicolumn{7}{l}{$(0,0,1)$}\\
 & 0 & B & $-9.0916$ & $0.00$ & $-9.0997$~\cite{Varga:1998ss} & $0.00$ \\
\hline
\multicolumn{7}{l}{$(1,0,0)$}\\
 & 0 & B & $-8.9862$ & $0.00$ & $-8.9876$~\cite{suzuki2000excited,matyus2012molecular} & $0.00$ \\
\hline
\multicolumn{7}{l}{$(1,0,1)$}\\
 & 0 & R & $-8.6223$ & $25.63$ & $-8.6532$~\cite{suzuki2004stochastic} & $27.21$ \\[2pt]
$\bm{\mathrm{Ps}(1)+\mathrm{Ps}(2)}$ & & & $\bm{-8.5036}$ & & & \\[2pt]
\multicolumn{7}{l}{$(0,0,0)$}\\
 & 0 & R & $-7.9413$ & $56.11$ & $-7.9376$~\cite{matyus2013resonances,ho1989resonant} & $68.03$ \\
 & 1 & R & $-7.8823$ & $0.93$ & $-7.8856$~\cite{matyus2013resonances} & $2.10$ \\[2pt]
$\bm{\mathrm{Ps}(1)+\mathrm{Ps}(3)}$ & & & $\bm{-7.5587}$ & & & \\
\end{tabular}
\end{ruledtabular}
\end{table}

%\subsubsection{$\mu^+\mu^+e^-e^-$}\label{subsubsec:mmEE}

\subsection{$\mu^+\mu^+e^-e^-$}\label{subsubsec:mmEE}

\begin{figure*}[t]
\centering
\includegraphics[width=\textwidth]{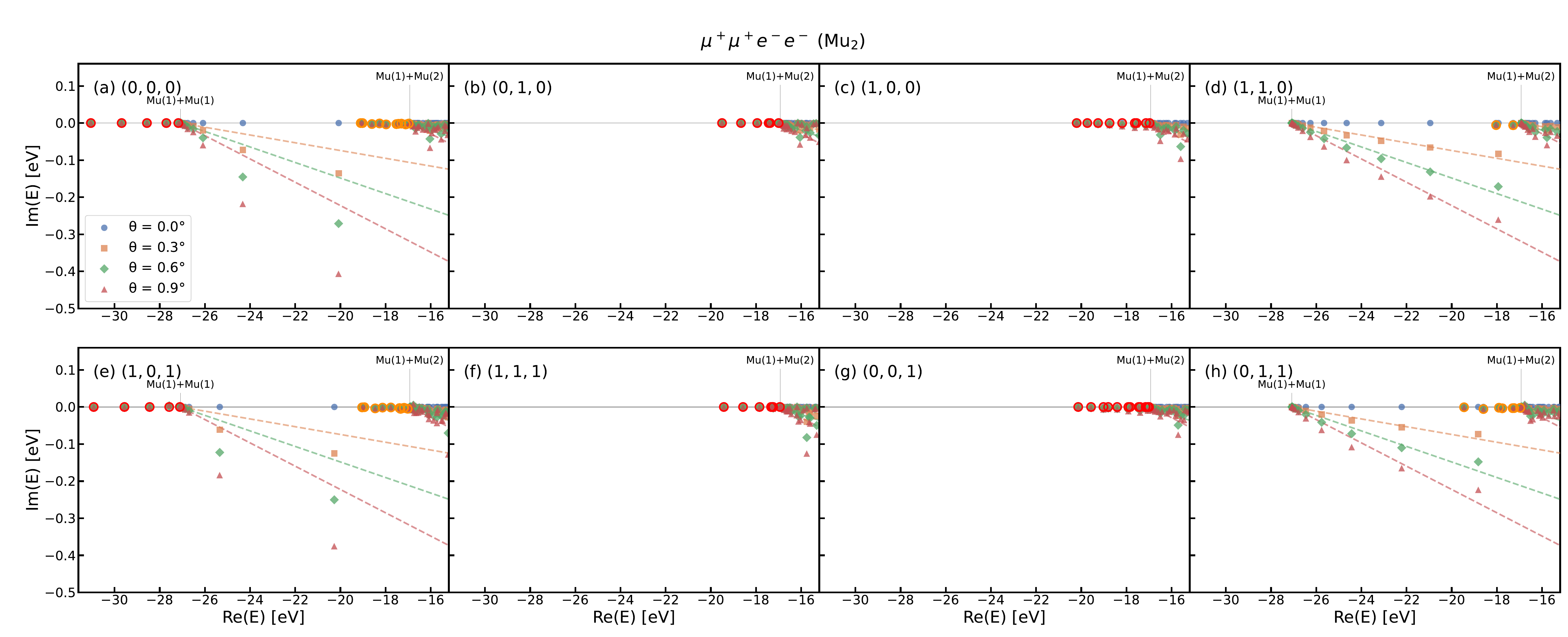}
\caption{\label{fig:mmEE} Complex-energy spectra of the $\mu^+\mu^+e^-e^-$
system under complex scaling for small angles $\theta=0,0.3,0.6,0.9^\circ$, for the
eight channels $(S_{\mu\mu},S_{ee},L)$: top row $L=0$ [(a)~$(0,0,0)$,
(b)~$(0,1,0)$, (c)~$(1,0,0)$, (d)~$(1,1,0)$] and bottom row $L=1$
[(e)~$(1,0,1)$, (f)~$(1,1,1)$, (g)~$(0,0,1)$, (h)~$(0,1,1)$], with each
column pairing the $S$- and $P$-wave channels of the same Born--Oppenheimer
electronic configuration. Continuum eigenvalues rotate along
$\mathrm{Arg}(E-E_{\rm th})=-2\theta$ from each open threshold. Discrete states
below the next threshold remain fixed. Bound states are shown in red and
Feshbach resonances in orange. The $\mathrm{Mu}(1)+\mathrm{Mu}(1)$ threshold
appears only in Pauli-allowed channels.}
\end{figure*}

The muonium--muonium molecule $\mu^+\mu^+e^-e^-$ is a
mass-asymmetric analogue of $\mathrm{Ps}_2$, with two heavy positive centers and
two light electrons. We label each channel by $(S_{\mu\mu},S_{ee},L)$, where
$S_{\mu\mu}$ and $S_{ee}$ are the spins of the $\mu^+\mu^+$ and $e^-e^-$ pairs.
The dissociation channels are $\mathrm{Mu}(n)+\mathrm{Mu}(n')$, with threshold
energies given by the sums of the corresponding muonium energy in
Table~\ref{tab:twobody}.  The lower threshold
contains two identical $\mathrm{Mu}(1S)$ atoms. The same exchange symmetry selection rule as
in $\mathrm{Ps}_2$ applies. This threshold is open only when
$S_{\mu\mu}+S_{ee}+L$ is even. Figure~\ref{fig:mmEE} shows the complex-energy
spectra of the eight spin--orbital channels.

In the Born--Oppenheimer picture, the light electrons
generate electronic potential curves for the $\mu^+$--$\mu^+$ motion. On a given
electronic potential curve, the exchange symmetry of the two $\mu^+$ ties the
muon-pair spin to the parity of the $\mu^+$--$\mu^+$ rotational wave function.
Thus, each electronic configuration is associated with one $S$-wave channel and one
$P$-wave channel. In the
$(S_{\mu\mu},S_{ee},L)$ notation, the configurations are labeled by the usual
molecular notation ${}^{2S_{ee}+1}\Lambda_{g/u}^{\pm}$. The four configurations
correlating to the two lowest thresholds are
\begin{align}\label{eq:mmEE_terms}
{}^1\Sigma_g^+&:\ (0,0,0)\ \text{and}\ (1,0,1),\notag\\
{}^3\Sigma_g^+&:\ (0,1,0)\ \text{and}\ (1,1,1),\notag\\
{}^1\Sigma_u^+&:\ (1,0,0)\ \text{and}\ (0,0,1),\\
{}^3\Sigma_u^+&:\ (1,1,0)\ \text{and}\ (0,1,1),\notag
\end{align}
mirroring the lowest states of the $\mathrm{H}_2$
molecule~\cite{matyus2012molecular}. This $S$--$P$ pairing within each configuration is
useful for comparing Fig.~\ref{fig:mmEE} with Table~\ref{tab:mmEE}.

The upper row of Fig.~\ref{fig:mmEE} contains the $S$-wave channels. The
$(0,0,0)$ channel corresponds to the bonding $^1\Sigma_g^+$ ground configuration.
In this channel we find five bound states ($v=0$--$4$), from $-31.05$ to
$-27.18~\mathrm{eV}$. Their energies agree with previous ECG
calculations~\cite{Wu:2024qan,Hu:2025cyj}. The $(1,1,0)$ channel corresponds to the $^3\Sigma_u^+$
configuration. For this channel, the
$\mathrm{Mu}(1)+\mathrm{Mu}(1)$ threshold is open, so the two listed states
($v=0,1$) are Feshbach resonances between the two lowest thresholds. In the $(1,0,0)$ ($^1\Sigma_u^+$) and
$(0,1,0)$ ($^3\Sigma_g^+$) channels the $\mathrm{Mu}(1)+\mathrm{Mu}(1)$
threshold is Pauli-forbidden, so the listed states are genuine bound states
below $\mathrm{Mu}(1)+\mathrm{Mu}(2)$.

The lower row of Fig.~\ref{fig:mmEE} contains the corresponding $P$-wave
 partners of these $S$-wave channels. The bonding $^1\Sigma_g^+$ configuration
connects $(0,0,0)$ with $(1,0,1)$. The $(1,0,1)$ channel has five bound states
below $\mathrm{Mu}(1)+\mathrm{Mu}(1)$ and 10 resonances ($v=0$-$9$)
between the two lowest thresholds. The $^3\Sigma_u^+$ configuration connects $(1,1,0)$
with $(0,1,1)$ channels. The $(0,1,1)$ channel is pauli-allowed, so it supports only
Feshbach resonances in this energy region. The $^1\Sigma_u^+$ configuration connects
$(1,0,0)$ with $(0,0,1)$ channels, while the $^3\Sigma_g^+$ configuration connects $(0,1,0)$ with
$(1,1,1)$ channels. In the $(0,0,1)$ and $(1,1,1)$ channels the lower threshold is
Pauli-forbidden, so all listed states are bound states below
$\mathrm{Mu}(1)+\mathrm{Mu}(2)$. As in $\mathrm{Mu}_2^+$
(Sec.~\ref{subsubsec:mmE}), the widely separated $\mu^+$ and $e^-$ length scales
make the resonance half-widths unstable under the small rotation angles used
here. We therefore list only the real parts. The full state list is given in
Table~\ref{tab:mmEE}.

The $(0,0,0)$ and $(1,0,1)$ channels
associated with $^{1}\Sigma_g^+$ exhibit a clear Born--Oppenheimer organization.
Nearly every resolved bound or resonant state in one channel has a corresponding state
in the other. This pairing is much less evident between (1,1,0) and (0,1,1) channels. Thus, the Born--Oppenheimer
approximation does not provide a complete description of the resonance spectrum, even though the muon is approximately $207$
times heavier than the electron. A fully dynamical four-body calculation is still required.

\begin{table*}[t]
\centering
\footnotesize
\renewcommand{\arraystretch}{1.0}
\caption{\label{tab:mmEE} Bound and resonant states of $\mu^+\mu^+e^-e^-$
below $\mathrm{Mu}(1)+\mathrm{Mu}(2)$, grouped by
$(S_{\mu\mu},S_{ee},L)$. The left and right panels collect the $S$- and
$P$-wave channels. Energies $E$ are in eV, and $v$ labels the state index within
each channel. ``B''/``R'' denote bound/resonant states. The
$\mathrm{Mu}(1)+\mathrm{Mu}(1)$ threshold is shown only when Pauli allowed. In
those channels, states between the two thresholds are Feshbach resonances. When
the lower threshold is Pauli-forbidden, the listed states are bound. Resonance
half-widths are not listed because the small rotation angles do not resolve them
reliably. In the rightmost column, results from the literature for the
$(0,0,0)$ bound states are presented for comparison, obtained with explicitly
correlated Gaussians~\cite{Hu:2025cyj}.}
\begin{tabular*}{0.49\textwidth}[t]{@{\extracolsep{\fill}}lrccc}
\hline\hline & & & \multicolumn{1}{c}{This work} & \multicolumn{1}{c}{Lit.} \\
 & $v$ & type & \multicolumn{1}{c}{$E$} & \multicolumn{1}{c}{$E$} \\\hline
\multicolumn{5}{l}{$(0,0,0)$}\\
 & 0 & B & $-31.0473$ & $-31.0486$~\cite{Hu:2025cyj} \\
 & 1 & B & $-29.6849$ & $-29.6878$~\cite{Hu:2025cyj} \\
 & 2 & B & $-28.5653$ & $-28.5752$~\cite{Hu:2025cyj} \\
 & 3 & B & $-27.7066$ & $-27.7247$~\cite{Hu:2025cyj} \\
 & 4 & B & $-27.1759$ & $-27.1923$~\cite{Hu:2025cyj} \\
$\bm{\mathrm{Mu}(1){+}\mathrm{Mu}(1)}$ & & & $\bm{-27.0804}$ & \\
 & 0 & R & $-19.1005$ & --- \\
 & 1 & R & $-19.0195$ & --- \\
 & 2 & R & $-18.6050$ & --- \\
 & 3 & R & $-18.2692$ & --- \\
 & 4 & R & $-17.9772$ & --- \\
 & 5 & R & $-17.5209$ & --- \\
 & 6 & R & $-17.4338$ & --- \\
 & 7 & R & $-17.3014$ & --- \\
 & 8 & R & $-17.1108$ & --- \\
 & 9 & R & $-16.9770$ & --- \\
\hline
\multicolumn{5}{l}{$(0,1,0)$}\\
 & 0 & B & $-19.5046$ & --- \\
 & 1 & B & $-18.6661$ & --- \\
 & 2 & B & $-17.9468$ & --- \\
 & 3 & B & $-17.4411$ & --- \\
 & 4 & B & $-17.3722$ & --- \\
 & 5 & B & $-16.9975$ & --- \\
\hline
\multicolumn{5}{l}{$(1,0,0)$}\\
 & 0 & B & $-20.2040$ & --- \\
 & 1 & B & $-19.7324$ & --- \\
 & 2 & B & $-19.2587$ & --- \\
 & 3 & B & $-18.7470$ & --- \\
 & 4 & B & $-18.1908$ & --- \\
 & 5 & B & $-17.6312$ & --- \\
 & 6 & B & $-17.5510$ & --- \\
 & 7 & B & $-17.1344$ & --- \\
 & 8 & B & $-16.9758$ & --- \\
\hline
\multicolumn{5}{l}{$(1,1,0)$}\\
 & 0 & R & $-18.0352$ & --- \\
 & 1 & R & $-17.2810$ & --- \\
$\bm{\mathrm{Mu}(1){+}\mathrm{Mu}(2)}$ & & & $\bm{-16.9253}$ & --- \\
\hline\hline\end{tabular*}
\hfill
\begin{tabular*}{0.49\textwidth}[t]{@{\extracolsep{\fill}}lrcc}
\hline\hline & & & \multicolumn{1}{c}{This work} \\
 & $v$ & type & \multicolumn{1}{c}{$E$} \\\hline
\multicolumn{4}{l}{$(1,0,1)$}\\
 & 0 & B & $-30.9226$ \\
 & 1 & B & $-29.5628$ \\
 & 2 & B & $-28.4449$ \\
 & 3 & B & $-27.5792$ \\
 & 4 & B & $-27.1074$ \\
$\bm{\mathrm{Mu}(1){+}\mathrm{Mu}(1)}$ & & & $\bm{-27.0804}$ \\
 & 0 & R & $-19.0393$ \\
 & 1 & R & $-18.9401$ \\
 & 2 & R & $-18.4613$ \\
 & 3 & R & $-18.1359$ \\
 & 4 & R & $-17.7670$ \\
 & 5 & R & $-17.3853$ \\
 & 6 & R & $-17.3371$ \\
 & 7 & R & $-17.1994$ \\
 & 8 & R & $-17.1776$ \\
 & 9 & R & $-17.0242$ \\
\hline
\multicolumn{4}{l}{$(1,1,1)$}\\
 & 0 & B & $-19.4260$ \\
 & 1 & B & $-18.5753$ \\
 & 2 & B & $-17.8470$ \\
 & 3 & B & $-17.3343$ \\
 & 4 & B & $-17.2653$ \\
 & 5 & B & $-17.2418$ \\
 & 6 & B & $-16.9432$ \\
\hline
\multicolumn{4}{l}{$(0,0,1)$}\\
 & 0 & B & $-20.1366$ \\
 & 1 & B & $-19.5687$ \\
 & 2 & B & $-19.0203$ \\
 & 3 & B & $-18.8104$ \\
 & 4 & B & $-18.4107$ \\
 & 5 & B & $-17.9173$ \\
 & 6 & B & $-17.8370$ \\
 & 7 & B & $-17.4605$ \\
 & 8 & B & $-17.3972$ \\
 & 9 & B & $-17.1546$ \\
 & 10 & B & $-17.0538$ \\
 & 11 & B & $-17.0012$ \\
\hline
\multicolumn{4}{l}{$(0,1,1)$}\\
 & 0 & R & $-19.4547$ \\
 & 1 & R & $-18.5970$ \\
 & 2 & R & $-17.9086$ \\
 & 3 & R & $-17.7584$ \\
 & 4 & R & $-17.3034$ \\
 & 5 & R & $-17.2367$ \\
 & 6 & R & $-16.9695$ \\
$\bm{\mathrm{Mu}(1){+}\mathrm{Mu}(2)}$ & & & $\bm{-16.9253}$ \\
\hline\hline\end{tabular*}\ \null% the trailing "\ \null" reproduces the exact panel spacing of the former \input{...} layout
\end{table*}

\section{Summary and Discussion}\label{sec:sum}

We have computed the bound and resonant spectra of three trilepton systems,
$e^-e^-e^+$, $\mu^+e^-e^-$, and $\mu^+\mu^+e^-$, and two tetralepton systems,
$e^+e^+e^-e^-$ and $\mu^+\mu^+e^-e^-$. The calculation covers the
natural-parity $S$- and $P$-wave channels and treats bound states and resonances
on the same footing. The complex-scaling method is used to identify the resonances.

The equal-mass $\mathrm{Ps}^-$ system has no heavy--light
separation, whereas $\mathrm{Mu}^-$ and $\mathrm{Mu}_2^+$ are conventionally
viewed as atomic and molecular systems, respectively. The occurrence of Gailitis--Damburg
resonance sequences in all three systems reveals universal near-threshold
dynamics that is not tied to these conventional structural classifications.
This universality originates from $2S$--$2P$ Stark mixing. When the resulting
attractive inverse-square interaction is supercritical, a geometric
accumulation of resonances occurs. The resonance density depends strongly on the mass configuration.
$\mathrm{Ps}^-$ and $\mathrm{Mu}^-$ contain only a few resolvable members,
whereas $\mathrm{Mu}_2^+$ exhibits much denser sequences, including 20 states
in the $^3P^o$ channel. Its fitted dipole parameters remain below their
asymptotic values, showing that the resolved spectrum have not yet reached the asymptotic regime.

% The deeper spectra display a molecular organization when a heavy--light separation of scales is present.
% In $\mathrm{Mu}_2^+$, the bonding $^2\Sigma_g^+$ Born--Oppenheimer curve carries the $^1S^e$ and $^3P^o$ levels,
% whereas the antibonding $^2\Sigma_u^+$ curve carries the $^3S^e$ and $^1P^o$ levels.
% The four-body $\mathrm{Mu}_2$ system shows an analogous molecular organization, with exchange symmetry pairing its $S$- and $P$-wave channels.
% For equal-mass $\mathrm{Ps}_2$, a controlled Born--Oppenheimer separation is unavailable, and the molecular terms provide only a symmetry classification.
% The calculation gives the $\mathrm{Ps}_2$ ground state at $-14.0410~\mathrm{eV}$ and finds bound and Feshbach states in both partial waves,
% including the $L=1$ state accessed by optical spectroscopy~\cite{Cassidy_2012}.

The $\mathrm{Mu}_2^+$ spectrum also displays a crossover between molecular and
atomic organizations. Its deeper bound and resonant states retain
Born--Oppenheimer-like channel pairings. Near the $n=2$ threshold, the atomic
$\mathrm{Mu}(2)+\mu^+$ structure become
dominant. The appropriate structural description
therefore changes with the energy and spatial scale of the states.

The tetralepton spectra further illustrate the roles of threshold degeneracy
and fermion exchange symmetry. In both systems, exchange symmetry makes the lowest
threshold Pauli-allowed in some channels and Pauli-forbidden in others. This
selection rule determines whether states below the next threshold are
bound or resonant. In $\mathrm{Ps}_2$, additional basis tests show that the
$(1,0,0)$ bound state is recovered only when the degenerate
$\mathrm{Ps}(1S)+\mathrm{Ps}(2S)$ and
$\mathrm{Ps}(1S)+\mathrm{Ps}(2P)$ configurations are included together.
% The trilepton spectra near the degenerate $n=2$ thresholds show the
% Gailitis--Damburg accumulation pattern. In this region, the spectator lepton
% couples the degenerate $2S$ and $2P$ states of the neutral atom through the
% linear Stark interaction. The induced charge--dipole potential has an attractive
% $1/R^2$ tail. Once this attraction is supercritical, the near-threshold levels
% accumulate geometrically. We characterize the resolved sequences by fitting
% $\ln|E_v-E_{\rm th}|$ as a function of the state index $v$. The resulting
% slopes define effective dipole parameters $s$ over the resolved energy range.
% The value of $s$ increases with the mass ratio $\eta=\mu_R/\mu_X$. The
% $\mu^+\mu^+e^-$ system therefore has the densest sequence, with
% $\eta\simeq104$.
In $\mathrm{Mu}_2$, the channels associated with $^1\Sigma_g^+$ BO configuration exhibit a
nearly one-to-one correspondence across the resolved bound and resonant
states. No such pairing occurs in the $^3\Sigma_u^+$ BO configuration. The Born--Oppenheimer organization is thus channel dependent.
Even a heavy-to-light mass ratio of approximately $207$, the BO approximation is insufficient to
fully describe the resonance spectrum.

% The Born--Oppenheimer and Gailitis--Damburg descriptions therefore characterize complementary dynamical regimes. This distinction is clearest in $\mathrm{Mu}_2^+$: its deeper bound and resonant states follow a molecular Born--Oppenheimer organization, whereas its near-threshold spectrum is governed by the atomic $\mathrm{Mu}(2)+\mu^+$ channel and the long-range charge--dipole interaction. Atomic and molecular structures are thus not fixed classifications determined solely by the constituent masses, but different organizations that emerge with energy and length scale.
The spectra reported here provide nonrelativistic benchmarks for few-body
resonance calculations. The results show how universal
threshold dynamics, mass imbalance, and exchange symmetry jointly shape
few-body spectra. Further developments in high-intensity positron and muon facilities may enable experimental searches
for these states.
% ~\cite{Achasov:2023gey,Prokscha:2008zz,Kanda:2023gqp,Bai:2024skk,
% Chen:2026tdg,Cai:2023caf,An:2025lws,Liu:2025ejy}.

% Together with the molecular bound and resonant states discussed above, these
% near-threshold sequences provide concrete targets for future few-lepton studies.
% High-intensity positron and muon facilities provide the experimental motivation
% for such studies~\cite{Achasov:2023gey,Prokscha:2008zz,Kanda:2023gqp,Bai:2024skk,Chen:2026tdg,Cai:2023caf,An:2025lws,Liu:2025ejy}. The
% present spectra provide a nonrelativistic QED reference and benchmarks for
% few-body resonance methods and molecular particle--antiparticle complexes.

\appendix
\section{Scaled spectra of $\mu^+\mu^+\mu^-$ and $\mu^+\mu^+\mu^-\mu^-$}\label{app:mmm}

The $\mu^+\mu^+\mu^-$ system is the equal-mass muonic analogue of
$e^-e^-e^+$. With only the constituent mass setting the scale, its
spectrum follows exactly from the $e^-e^-e^+$ results by the rescaling
$E\to(m_\mu/m_e)\,E$ ($m_\mu/m_e=206.768$). The thresholds are those of true
muonium ($\mathrm{TM}=\mu^+\mu^-$). The scaled energy levels are listed in
Table~\ref{tab:mmm}.

\begin{table}[htbp]
\renewcommand{\arraystretch}{1.3}
\centering
\caption{\label{tab:mmm} Scaled spectrum of $\mu^+\mu^+\mu^-$, obtained from
Table~\ref{tab:eee} by $E\to(m_\mu/m_e)E$. ``B''/``R'' denote bound/resonant
states. $\mathrm{TM}=\mu^+\mu^-$ denotes true muonium.}
\begin{ruledtabular}
\begin{tabular}{l c c c c}
 & $v$ & type & $E$ (eV) & $\Gamma/2$ (meV) \\
\hline
\multicolumn{5}{l}{$^1S^e\ (0,0)$}\\
 & 0 & B & $-1474.16$ & $0.00$      \\[2pt]
$\bm{\mathrm{TM}(1)+\mu^+}$ & & & $\bm{-1406.61}$ & \\[2pt]
 & 0 & R & $-427.78$  & $119.10$ \\
 & 1 & R & $-358.12$  & $25.02$  \\
 & 2 & R & $-352.17$  & $31.22$  \\
\hline
\multicolumn{5}{l}{$^3S^e\ (1,0)$}\\
 & 0 & R & $-357.49$  & $0.01$   \\
 & 1 & R & $-352.07$  & $54.17$  \\
\hline
\multicolumn{5}{l}{$^1P^o\ (0,1)$}\\
 & 0 & R & $-355.34$  & $2.69$   \\
 & 1 & R & $-351.82$  & $20.06$  \\
\hline
\multicolumn{5}{l}{$^3P^o\ (1,1)$}\\
 & 0 & R & $-412.58$  & $365.98$ \\
 & 1 & R & $-355.01$  & $46.73$  \\
 & 2 & R & $-351.79$  & $21.50$  \\[2pt]
$\bm{\mathrm{TM}(2)+\mu^+}$ & & & $\bm{-351.65}$ & \\[2pt]
\end{tabular}
\end{ruledtabular}
\end{table}

The equal-mass tetralepton $\mu^+\mu^+\mu^-\mu^-$ follows in the same way from
the $e^+e^+e^-e^-$ results (Table~\ref{tab:eeEE}), now with the
true-muonium molecule setting the thresholds. The scaled energy levels are listed in
Table~\ref{tab:mmmm}.

\begin{table}[htbp]
\renewcommand{\arraystretch}{1.3}
\centering
\caption{\label{tab:mmmm} Scaled spectrum of $\mu^+\mu^+\mu^-\mu^-$, obtained
from Table~\ref{tab:eeEE} by $E\to(m_\mu/m_e)E$. Channels are labeled by
$(S_{12},S_{34},L)$, and $\mathrm{TM}=\mu^+\mu^-$ denotes true muonium.}
\begin{ruledtabular}
\begin{tabular}{l c c c c}
 & $v$ & type & $E$ (eV) & $\Gamma/2$ (eV) \\
\hline
\multicolumn{5}{l}{$(0,0,0)$}\\
 & 0 & B & $-2903.23$ & $0.00$ \\[2pt]
$\bm{\mathrm{TM}(1)+\mathrm{TM}(1)}$ & & & $\bm{-2813.22}$ & \\[2pt]
 & 0 & R & $-1851.81$ & $16.07$ \\
 & 1 & B & $-1770.20$ & $0.00$  \\
\hline
\multicolumn{5}{l}{$(0,0,1)$}\\
 & 0 & B & $-1879.85$ & $0.00$ \\
\hline
\multicolumn{5}{l}{$(1,0,0)$}\\
 & 0 & B & $-1858.06$ & $0.00$ \\
\hline
\multicolumn{5}{l}{$(1,0,1)$}\\
 & 0 & R & $-1782.82$ & $5.30$ \\[2pt]
$\bm{\mathrm{TM}(1)+\mathrm{TM}(2)}$ & & & $\bm{-1758.27}$ & \\[2pt]
\multicolumn{5}{l}{$(0,0,0)$}\\
 & 0 & R & $-1642.01$ & $11.60$ \\
 & 1 & R & $-1629.81$ & $0.19$  \\[2pt]
$\bm{\mathrm{TM}(1)+\mathrm{TM}(3)}$ & & & $\bm{-1562.90}$ & \\
\end{tabular}
\end{ruledtabular}
\end{table}

\section{Angular matrix for the Gailitis--Damburg dipole interaction}\label{app:GD_angular}

The angular factor in Eq.~\eqref{eq:GDD} is evaluated in the spherical-tensor
basis. We use
$C_q^{(1)}(\hat{\bm r})=\sqrt{4\pi/3}\,Y_{1q}(\hat{\bm r})$ and the scalar-product
convention
\begin{equation}
C^{(1)}(\hat{\bm r})\!\cdot\!C^{(1)}(\hat{\bm R})
=\sum_q(-1)^q C_q^{(1)}(\hat{\bm r})C_{-q}^{(1)}(\hat{\bm R}).
\end{equation}
With this convention, $\hat{\bm r}\!\cdot\!\hat{\bm R}
=C^{(1)}(\hat{\bm r})\!\cdot\!C^{(1)}(\hat{\bm R})$, and
\begin{equation}\label{eq:GDangular}
\begin{split}
A_{ij}={}&(-1)^{L+\ell_j+\lambda_i}
\begin{Bmatrix}
\ell_i&\lambda_i&L\\
\lambda_j&\ell_j&1
\end{Bmatrix}  \\
&\times
\langle\ell_i\|C^{(1)}\|\ell_j\rangle
\langle\lambda_i\|C^{(1)}\|\lambda_j\rangle .
\end{split}
\end{equation}
The reduced matrix elements follow the Edmonds convention~\cite{edmonds1957},
\begin{equation}
\langle\ell\|C^{(1)}\|\ell'\rangle
=(-1)^\ell\sqrt{(2\ell+1)(2\ell'+1)}
\begin{pmatrix}\ell&1&\ell'\\0&0&0\end{pmatrix}.
\end{equation}

\section{Basis parameters}\label{app:Gaussian_par}

This appendix lists the basis parameters used in the ESVM calculation of
Sec.~\ref{subsec:ESVM}. In the random ECG basis,
the width parameters are sampled from the listed radial windows. In the
molecular basis, the inter-cluster widths follow the geometric progression of
Eq.~\eqref{eq:geometric}. The intra-cluster factor is the two-body atom optimized
up to the listed $4S$ target. For $L=1$, the global-vector weights are written as
a central value plus a normal fluctuation, $\mathcal N(\mathbf
0,\boldsymbol\sigma)$.

\begin{widetext}
\begingroup\scriptsize
\setlength{\arraycolsep}{2pt}
\paragraph{$e^-e^-e^+$.}
\[
\begin{array}{@{}l@{\ }l@{}}
\mathrm{random}\left\{\begin{array}{ll}
r_{e^-e^-} &\in [0.005,10]\,\mathrm{nm}\\
r_{e^-e^+} &\in [0.005,10]\,\mathrm{nm}\\
N_{\max} &=6000\ (L=0),\ 10000\ (L=1)\\
\left[u_{e^-e^-},u_{(e^-e^-)-e^+}\right]&\in (1,0.5)+\mathcal N(\mathbf 0,(0.6,0.6))
\end{array}\right.
&
\mathrm{molecular}\left\{\begin{array}{ll}
r_{\rm inter}=r_{\mathrm{Ps}-e^-} &\in [0.005,30]\,\mathrm{nm},\ n_{\max}=35\\
r_{\rm intra}=r_{e^-e^+} &\in [0.0002,20]\,\mathrm{nm},\ n_{\max}=20\\
\left[u_{e^-e^+},u_{\mathrm{Ps}-e^-}\right]&\in \begin{cases}(1,0)\\ (0,1)\\ (0.5,0.5)\end{cases}
\end{array}\right.
\end{array}
\]

\paragraph{$\mu^+e^-e^-$.}
\[
\begin{array}{@{}l@{\ }l@{}}
\mathrm{random}\left\{\begin{array}{ll}
r_{e^-e^-} &\in [0.0025,5]\,\mathrm{nm}\\
r_{\mu^+e^-} &\in [0.0025,5]\,\mathrm{nm}\\
N_{\max} &=6000\ (L=0),\ 10000\ (L=1)\\
\left[u_{e^-e^-},u_{(e^-e^-)-\mu^+}\right]&\in (1,0.5)+\mathcal N(\mathbf 0,(0.6,0.6))
\end{array}\right.
&
\mathrm{molecular}\left\{\begin{array}{ll}
r_{\rm inter}=r_{\mathrm{Mu}-e^-} &\in [0.0025,15]\,\mathrm{nm},\ n_{\max}=35\\
r_{\rm intra}=r_{\mu^+e^-} &\in [0.0001,10]\,\mathrm{nm},\ n_{\max}=20\\
\left[u_{\mu^+e^-},u_{\mathrm{Mu}-e^-}\right]&\in \begin{cases}(1,0)\\ (0,1)\\ (0.5,0.5)\end{cases}
\end{array}\right.
\end{array}
\]

\paragraph{$\mu^+\mu^+e^-$.}
\[
\begin{array}{@{}l@{\ }l@{}}
\mathrm{random}\left\{\begin{array}{ll}
r_{\mu^+\mu^+} &\in [0.005,25]\,\mathrm{nm}\\
r_{\mu^+e^-} &\in [0.005,25]\,\mathrm{nm}\\
N_{\max} &=5300\ (L=0),\ 10000\ (L=1)\\
\left[u_{\mu^+\mu^+},u_{(\mu^+\mu^+)-e^-}\right]&\in (1,0.5)+\mathcal N(\mathbf 0,(0.6,0.6))
\end{array}\right.
&
\mathrm{molecular}\left\{\begin{array}{ll}
r_{\rm inter}=r_{\mathrm{Mu}-\mu^+} &\in [0.0025,15]\,\mathrm{nm},\ n_{\max}=35\\
r_{\rm intra}=r_{\mu^+e^-} &\in [0.0001,10]\,\mathrm{nm},\ n_{\max}=20\\
\left[u_{\mu^+e^-},u_{\mathrm{Mu}-\mu^+}\right]&\in \begin{cases}(1,0)\\ (0,1)\\ (0.5,0.5)\end{cases}
\end{array}\right.
\end{array}
\]

\paragraph{$e^+e^+e^-e^-$.}
\[
\begin{array}{@{}l@{\ }l@{}}
\mathrm{random}\left\{\begin{array}{ll}
r_{e^-e^-} &\in [0.008,4]\,\mathrm{nm}\\
r_{e^+e^+} &\in [0.008,4]\,\mathrm{nm}\\
r_{e^-e^+} &\in [0.008,4]\,\mathrm{nm}\\
N_{\max} &=10000\\
\left[u_{e^-e^-},u_{e^+e^+},u_{(e^-e^-)-(e^+e^+)}\right]&\in (0.5,0.3,1)+\\
&\quad\mathcal N(\mathbf 0,(0.6,0.6,0.9))
\end{array}\right.
&
\mathrm{molecular}\left\{\begin{array}{ll}
r_{\rm inter}=r_{\mathrm{Ps}-\mathrm{Ps}}\,(L{=}0) &\in \begin{cases}[0.02,1]\,\mathrm{nm},\ n_{\max}=20\\ [1,2.8]\,\mathrm{nm},\ n_{\max}=15\end{cases}\\
r_{\rm inter}=r_{\mathrm{Ps}-\mathrm{Ps}}\,(L{=}1) &\in [0.02,1.4]\,\mathrm{nm},\ n_{\max}=20\\
r_{\rm intra}=r_{e^-e^+} &\in [0.008,20]\,\mathrm{nm},\ n_{\max}=15\\
\left[u_{(e^-e^+)_1},u_{(e^-e^+)_2},u_{\mathrm{Ps}-\mathrm{Ps}}\right]&\in \begin{cases}(1,0,0)\\ (0,0,1)\\ (0.5,0,0.5)\end{cases}
\end{array}\right.
\end{array}
\]

\paragraph{$\mu^+\mu^+e^-e^-$.}
\[
\begin{array}{@{}l@{\ }l@{}}
\mathrm{random}\left\{\begin{array}{ll}
r_{\mu^+\mu^+} &\in [0.012,2]\,\mathrm{nm}\\
r_{e^-e^-} &\in [0.012,2]\,\mathrm{nm}\\
r_{\mu^+e^-} &\in [0.012,2]\,\mathrm{nm}\\
N_{\max} &=10000\\
\left[u_{\mu^+\mu^+},u_{e^-e^-},u_{(\mu^+\mu^+)-(e^-e^-)}\right]&\in (0.5,0.3,1)+\\
&\quad\mathcal N(\mathbf 0,(0.6,0.6,0.9))
\end{array}\right.
&
\mathrm{molecular}\left\{\begin{array}{ll}
r_{\rm inter}=r_{\mathrm{Mu}-\mathrm{Mu}}\,(L{=}0) &\in \begin{cases}[0.04,1]\,\mathrm{nm},\ n_{\max}=20\\ [1,3.7]\,\mathrm{nm},\ n_{\max}=15\end{cases}\\
r_{\rm inter}=r_{\mathrm{Mu}-\mathrm{Mu}}\,(L{=}1) &\in [0.04,3.7]\,\mathrm{nm},\ n_{\max}=20\\
r_{\rm intra}=r_{\mu^+e^-} &\in [0.004,10]\,\mathrm{nm},\ n_{\max}=15\\
\left[u_{(\mu^+e^-)_1},u_{(\mu^+e^-)_2},u_{\mathrm{Mu}-\mathrm{Mu}}\right]&\in \begin{cases}(0.5,0,0.5)\\ +\mathcal N(\mathbf 0,(0.3,0.3,0.3))\\ (0.5,1,0.5)\\ +\mathcal N(\mathbf 0,(0.3,0.3,0.3))\end{cases}
\end{array}\right.
\end{array}
\]
\endgroup
\end{widetext}

The muonic systems $\mu^+\mu^+\mu^-$ and $\mu^+\mu^+\mu^-\mu^-$ are obtained
by mass rescaling the equal-mass electron systems and require no separate basis.

\begin{acknowledgments}
We are grateful to Wei-Lin Wu, Yan-Ke Chen, and Xin-He Zheng for helpful discussions. This project was supported by the
National Natural Science Foundation of China (Grant No.~12475137), and ERC
NuclearTheory (Grant No.~885150). Y. M. is supported by the Alexander von Humboldt Foundation. The computational resources were supported by
the High-performance Computing Platform of Peking University.
\end{acknowledgments}

\bibliography{Ref}
\end{document}